\pdfoutput=1
\documentclass[fleqn,usenatbib]{mnras}

\usepackage[T1]{fontenc}
\usepackage{ae,aecompl}

\usepackage{soul,color,xcolor}
\usepackage{hyperref}
\usepackage{graphicx}
\usepackage{amsmath}
\usepackage{amssymb}
\usepackage{lineno}
\usepackage{lastpage}
\usepackage[absolute,overlay]{textpos}
\usepackage{newtxtext,newtxmath}

\newcommand{\md}{\mathrm{d}}
\definecolor{cyan}{rgb}{0.0, 1.0, 1.0}

\title[Serendipitous Kilonovae]{Serendipitous Discoveries of Kilonovae in the LSST Main Survey: Maximising Detections of Sub-Threshold Gravitational Wave Events}

\author[C. N. Setzer et al.]{Christian N. Setzer,$^{1,2}$\thanks{E-mail: christian.setzer@fysik.su.se}
Rahul Biswas,$^{1,2}$
Hiranya V. Peiris,$^{1,2,3}$
\newauthor Stephan Rosswog,$^{1,4}$
Oleg Korobkin,$^{5,6,7}$
and Ryan T. Wollaeger$^{6}$
\newauthor \large (The LSST Dark Energy Science Collaboration) \vspace{2.5pt}
\\
$^{1}$The Oskar Klein Centre for Cosmoparticle Physics, Stockholm University, AlbaNova, Stockholm SE-106 91, Sweden\\
$^{2}$Department of Physics, Stockholm University, AlbaNova, Stockholm SE-106 91, Sweden \\
$^{3}$Department of Physics and Astronomy, University College London, Gower Street, London WC1E 6BT, UK\\
$^{4}$Department of Astronomy, Stockholm University, AlbaNova, Stockholm SE-106 91, Sweden\\
$^{5}$CCS-7, Los Alamos National Laboratory, P.O. Box 1663, Los Alamos, NM 87545, USA\\
$^{6}$Center for Theoretical Astrophysics, Los Alamos National Laboratory, Los Alamos, NM 87545, USA\\
$^{7}$Joint Institute for Nuclear Astrophysics - Center for the Evolution of the Elements, USA
}

\date{Accepted XXX. Received YYY; in original form ZZZ}

\pubyear{2019}

\begin{document}
\label{firstpage}
\pagerange{\pageref{firstpage}--\pageref{LastPage}}
\maketitle

\begin{textblock*}{5cm}(16.3cm,2cm)
  \mbox{LA-UR-18-31553}
\end{textblock*}

\begin{abstract}
We investigate the ability of the Large Synoptic Survey Telescope (LSST) to discover kilonovae (kNe) from binary neutron star (BNS) and neutron star-black hole (NSBH) mergers, focusing on serendipitous detections in the Wide-Fast-Deep (WFD) survey. We simulate observations of kNe with proposed LSST survey strategies, focusing on cadence choices that are compatible with the broader LSST cosmology programme. If all kNe are identical to GW170817, we find the baseline survey strategy will yield 58 kNe over the survey lifetime. If we instead assume a representative population model of BNS kNe, we expect to detect only 27 kNe. However, we find the choice of survey strategy significantly impacts these numbers and can increase them to 254 kNe and 82 kNe over the survey lifetime, respectively. This improvement arises from an increased cadence of observations between different filters with respect to the baseline. We then consider the detectability of these BNS mergers by the Advanced LIGO/Virgo (ALV) detector network. If the optimal survey strategy is adopted, 202 of the GW170817-like kNe and 56 of the BNS population model kNe are detected with LSST but are below the threshold for detection by the ALV network. This represents, for both models, an increase by a factor greater than 4.5 in the number of detected sub-threshold events over the baseline strategy. These sub-threshold events would provide an opportunity to conduct electromagnetic-triggered searches for signals in gravitational-wave data and assess selection effects in measurements of the Hubble constant from standard sirens, e.g., viewing angle effects.
\end{abstract}

\begin{keywords}
surveys -- binaries: general -- stars: neutron -- stars: black holes -- gravitational waves -- cosmology: observations
\end{keywords}



\section{Introduction}
The Large Synoptic Survey Telescope (LSST)\footnote{\url{http://www.lsst.org}} will detect thousands of new transients on a nightly basis \citep{Ridgway2014}. This is expected to include many transients that occur rarely, such as the electromagnetic (EM) signals from the mergers of binary neutron stars (BNS) and neutron star black hole (NSBH) binaries. These EM signals are known as kilonovae (kNe). The coincident detection of gravitational waves (GW) and EM waves in 2017, an event designated GW170817, provided the first unambiguous evidence that kNe exist \citep{TheLIGOScientificCollaboration2017, Abbott2017a, Collaboration2017, Cowperthwaite2017, Kasliwal2017, Tanvir2017, Soares-Santos2017}. These signals were consistent with GW from a BNS merger accompanied by a short gamma-ray burst (GRB) jet and optical/infrared emission from {\it r}-process decay \citep{Collaboration2017}. This abundance of corroborating information confirmed long-standing predictions of multi-messenger signals from BNS mergers \citep{Kochanek1993, Li1998}. Even with one event a wealth of knowledge was gained; it was possible to place limits on the speed of GW \citep{Abbott2017b} and measure the Hubble constant using the GW signal and host galaxy redshift \citep{Schutz1986a}, yielding: $H_0 = 70.0^{+12.0}_{-8.0}\, \mathrm{km \, s^{-1}\,Mpc^{-1}}$ \citep{Abbott2017}.

Given the luminosity of GW170817 and the theoretical peak of the optical/infrared counterpart of compact object mergers, LSST is expected to detect kNe at greater distances than possible with contemporaneous GW-detector networks \citep{Chen2017a}. Detecting a population of kNe independent of the selection from GW or GRB triggers will provide a complementary determination of the rates of BNS and NSBH mergers and provide insight into the distribution of kNe ejecta masses, ejecta velocities, and opacities. Such a kNe population will enable studies of the connection between kNe, cosmic star-formation rates, and host-galaxy properties \citep{Vangioni2016, Blanchard2017}. In addition, such a population of serendipitously detected kNe will facilitate the study of the selection effects in measuring the Hubble constant from kNe identified via GW triggers \citep{Chen2017b, Mortlock2018}.

The EM detections of kNe can also be used as a trigger for searches of compact binary merger signals in archival GW data that otherwise would have fallen below the threshold to claim a detection \citep{Kochanek1993, Acernese2007, Kelley2012}. The time and position localisation, including sky position and distance information, lower the necessary threshold to reject false alarms for a GW signal detection \citep{Kelley2012}. With the redshift information from the EM signal and the luminosity distance from the GW measurements, these objects can be used as an independent probe of the Hubble constant \citep{Schutz1986a, Nissanke2013, Vitale2018}. This will yield a larger sample of standard sirens and accelerate efforts to resolve the Hubble constant tension \citep{Chen2018, Feeney2018, DiValentino2018, Mortlock2018}. Precise independent measurements of the Hubble constant are possible due to standard sirens having different systematics and modelling assumptions compared with other cosmological probes such as standard candles and the inverse distance ladder \citep{Holz2005, Dalal2006, Nissanke2010}. It may be possible, with a sample of as few as fifty events \citep{Feeney2018, Chen2018}, to discern if the standard siren measurements of the Hubble constant are consistent with current measurements from other sources \citep{Feeney2018}.

Before the start of LSST operations, Advanced LIGO and Advanced Virgo are expected to begin operating at design sensitivity. Additionally, Advanced LIGO will undergo substantial improvements to become LIGO A+ in the early to mid 2020s. Furthermore, KAGRA and LIGO India are expected to begin observations by 2025 \citep{TheKAGRACollaboration2013}. This rapid expansion of the GW-observatory network will improve localisation and sensitivity for detections of GW signals \citep{TheKAGRACollaboration2013}.

Target-of-Opportunity (ToO) observations are an efficient way to detect the EM signal from kNe triggered by detections of GW signals \citep{Margutti2018, Cowperthwaite2018}. An event provides a trigger for optical follow-up if it is detected by the GW-detector network with a signal-to-noise ratio (SNR) greater than some threshold; commonly an SNR of 12 is adopted for BNS mergers \citep{Huang2018}. Localisation from GW detections will improve to tens of degrees during the period of LSST \citep{Fairhurst2014}, which can be covered by LSST in only a few observations. \citet{Margutti2018} estimate that LSST will be able identify $1{-}10$ BNS kNe per year for an optimal ToO strategy comprising only two per cent of the survey time. This will enable population studies of BNS and NSBH mergers and facilitate detailed studies of the astrophysics of these objects \citep{Cowperthwaite2018, Margutti2018}.

Given the duty-cycles of GW observatories, this number of ToO kNe could be significantly modified. Historically, observing time comprises only $60{-}70$ per cent of a GW detector's operations due to considerable downtimes related to environmental disturbances and commissioning \citep{Biscans2018}. Long periods of inactivity for technical upgrades are also common and further reduce a GW detector's observing time. If even one detector is inoperative, this decreases the sensitivity of the detector network. The decrease in sensitivity leads to poorer sky localisation and a decreased distance to which the GW counterparts of kNe might be detected \citep{Schutz2011}. It is likely that several GW observatories will be offline during the operation of LSST, making ToO detections of kNe uncertain.

In the absence of ToO observations, a survey strategy could be designed specifically to detect kNe \citep{Andreoni2018}. A survey strategy could also be designed that changes behaviour to become more sensitive to detecting kNe during the operational periods of GW observatories. However, it is unlikely that the entire LSST WFD survey will be designed to meet only one science case or function differently based on the operational cycles of GW observatories. For these reasons, we aim to maximise the number of serendipitously detected kNe at all times, using a survey strategy that also fulfils the other science goals of LSST. We expect a survey strategy that maximises serendipitous detections will also lead to an increase in the number of detected kNe that are below the threshold for detection in GW-detector networks.

It should be noted that GW are not the only source of ToO triggers. GRBs can also act as triggers for optical follow-up \citep{Greiner1995}. Although they operate more continuously than GW observatories, detections with gamma-ray facilities are generally poorly localised \citep{Connaughton2014}, and also occur at rates lower than those expected for kNe due to the limited viewing angles for which short-GRB jets are observable \citep{Coward2011, Jin2018}. With serendipitous detections of kNe it will also be possible to study the selection effects of short GRBs and the relation between kNe and the short GRB population.

\citet{Rosswog2016a, Scolnic2017a, Cowperthwaite2018} have investigated the ability of LSST to serendipitously detect kNe. \citet{Rosswog2016a} used several kN models constructed using smoothed-particle hydrodynamics simulations and considered detectability based on the limiting magnitudes of future surveys. \citet{Scolnic2017a} considered the case of the event GW170817 as representative of all BNS kNe, while simulating observations for ten past, present, and future surveys including LSST, with a single realistic survey strategy for each. \citet{Cowperthwaite2018} considered detections of kNe using three LSST survey strategies and 27 kN model light curves from {\sc MOSFiT} \citep{Villar2017a, Guillochon2017}.

We build on the works of \citet{Rosswog2016a, Scolnic2017a} in several ways. We consider the ability of LSST to serendipitously detect kNe but with the eventual aim of using these detections as triggers to search catalogues of GW data to find sub-threshold GW events. The present goal is to determine which features of several proposed survey strategies detect the most kNe. We evaluate seven simulated LSST surveys including the current baseline strategy ({\sf kraken\_2026}, which has replaced {\sf minion\_1016} and {\sf baseline2018a}) and strategies that represent the major changes currently proposed to the LSST Wide-Fast-Deep (WFD) survey. Crucially, the strategies we consider are also consistent with the observational properties needed to meet the science requirements of the broad range of LSST cosmology probes \citep{LSSTScienceCollaboration2017, Lochner2018a, Scolnic2018}. For clarity, we adopt new descriptive names for the survey strategies considered, see \autoref{tab: cadence_names}. Further details about each survey strategy can be found in Sec.\ \ref{subsec:survey_strat}.

\begin{table}
  \centering
  \begin{tabular}{l|l}
    Designation & Descriptive name \\
    \hline
    {\sf kraken\_2026} & {\sf opsim\_baseline} \\
    {\sf alt\_sched\_rolling} & {\sf alt\_sched\_rolling} \\
    {\sf kraken\_2042} & {\sf opsim\_single\_exp} \\
    {\sf nexus\_2097} & {\sf opsim\_large\_rolling\_3yr} \\
    {\sf pontus\_2002} & {\sf opsim\_large} \\
    {\sf pontus\_2489} & {\sf opsim\_20s\_exp} \\
    {\sf pontus\_2573} & {\sf fbs\_mixed\_filter\_pairs} \\
    \hline
  \end{tabular}
  \caption{Simulation designations for LSST survey strategies mapped to the descriptive names used in this work.}
  \label{tab: cadence_names}
\end{table}

Following \citet{Scolnic2017a}, we consider the case where all BNS kNe are represented by the observed spectral energy distribution (SED) evolution of GW170817. We additionally consider a physically-motivated population model of BNS kNe and, for the first time, consider a population model of NSBH kNe. The use of a population model gives a wide parameter range to sample properties that generate kN light curves \citep{Rosswog2016a,Rosswog2018}. Though not yet observed, it is expected from theory that NSBH mergers will also be accompanied by similar EM and GW signals to that of BNS mergers \citep{Metzger2012}. The optical and near-infrared signals, in each case, are expected to come from dynamical ejecta, tidal tails unbound during merger, and post-merger accretion disk winds \citep{Rosswog2013}. Both of these components are powered by the decay of heavy {\it r}-process nuclei \citep{Rosswog2015, Fernandez2016}, and their EM signals evolve on the order of days, reaching absolute magnitudes of approximately $-16$ \citep{Barnes2013}. Indeed, emission indicating the presence of {\it r}-process in dynamical ejecta was seen in the light curves of GW170817 \citep{Smartt2017, Villar2017b, Cowperthwaite2017}.

We simulate observations using realistic rates and source distributions, and determine detections of kNe using multiple sets of criteria. For kNe that are detected, we simulate an approximation to their counterpart GW signals and calculate the associated SNRs for observations with the Advanced LIGO/Virgo (ALV) detector network. We perform this calculation to estimate the number of BNS mergers that LSST could detect that will not also be detected by GW observatories, and to investigate how this number changes in relation to the total number of serendipitously detected kNe.

In Sec.\ \ref{sec: modeling} we describe the different kN models used in our work and explain how we generate their light curves. This section also presents our methodology for calculating the associated GW signals and modelling the cosmological distribution of kNe. Section \ref{sec: sims} describes the simulation and analysis tools used to make mock observations of these kNe given realistic simulated survey strategies. In Sec.\ \ref{sec: criteria} we discuss the criteria used to determine the kNe observations that are considered detections. Section \ref{sec: results} presents the kN detection results from our simulated observations with different survey strategies. We also present the results of the GW SNR calculation: the number of kNe that are detected by LSST but not detected by ALV for each survey strategy and each BNS kN model. In Sec.\ \ref{sec: discussion} we discuss the impact of survey strategy choice on the detectability of kNe, and the prospects for optimisation of the LSST WFD survey strategy to maximise these detections. Finally, in Sec.\ \ref{sec: conclusion} we summarise our findings and present conclusions.

\section{Modelling}\label{sec: modeling}
\subsection{Kilonova Models}\label{sec:kne}
To simulate the EM signal from kNe we separately consider two models. The first model assumes the evolution of the kNe SEDs are identical to the SED evolution from an observed event. The only known observation of a kN is GW170817. We consider this as a BNS kN as the combined analysis of the EM and GW data for GW170817 has shown the event to be consistent with the merger of two neutron stars, disfavouring the case of a NSBH merger \citep{Hinderer2018}. To characterise this kN, we use the time-series SED provided by the Dark Energy Survey (DES) that was used in the analysis of \citet{Scolnic2017a}. This model uses multi-band photometry to calibrate an SED time-series to observations, using photometry from \citet{Soares-Santos2017} and \citet{Cowperthwaite2017}. Notably, this captures the early-blue and late-red kN components that have been described in literature \citep{Tanaka2017, Tanvir2017, Perego2017, Cowperthwaite2017}.

\begin{table}
  \centering
  \begin{tabular}{c|c|c|c}
    Parameter & Binary Type & Range & Units \\
    \hline
    $\kappa$ & BNS & $\mathrm{binomial}[1, 10]$ & $\mathrm{cm^2\, g^{-1}}$ \\
     & NSBH & $10$ & $\mathrm{cm^2\, g^{-1}}$ \\
    \hline
    ${m_\mathrm{ej}}$ & BNS & $[0.01, 0.2]$ & $\mathrm{M_{\odot}}$ \\
    & NSBH & $[0.05, 0.2]$ & $\mathrm{M_{\odot}}$ \\
    \hline
    ${v_\mathrm{ej}}$ & BNS & $[0.01, \, 0.5 ({m_\mathrm{ej}}/0.01)^{-\log_{20}(2)}]$ & $\mathrm{c}$ \\
     & NSBH & $[0.05, \, 0.5 ({m_\mathrm{ej}}/0.01)^{-\log_{20}(2)}]$ & $\mathrm{c}$\\
     \hline
  \end{tabular}
  \caption{The space of parameters describing the population of kNe generated by the SAEE (Population BNS/NSBH) model.}
  \label{tab: ross_params}
\end{table}

Although this is a reasonable first approximation, based on the diversity of other EM transients it would be naive to assume that GW170817 is representative of the EM properties of the entire range of neutron star mergers. Thus, we adopt an alternative model that is physically motivated and spans a parameter space describing a population of kNe. We have chosen a spherically symmetric semi-analytic eigenmode expansion (SAEE) approach for the SED evolution of kNe \citep{Wollaeger2018, Rosswog2018}. This model is based on the work of \citet{Pinto2000} and has been shown to faithfully approximate the results of full multi-group Monte Carlo radiative transfer calculations with the {\sc SuperNu} code \citep{Wollaeger2013} to a much greater accuracy than earlier such models, e.g., by \citet{Grossman2014}. In addition, the SAEE model is computationally efficient when compared to numerical radiative transfer simulations. This enables rapid exploration of the full parameter space of BNS and NSBH kNe.

\begin{figure}
  \centering
  \includegraphics[width=\columnwidth]{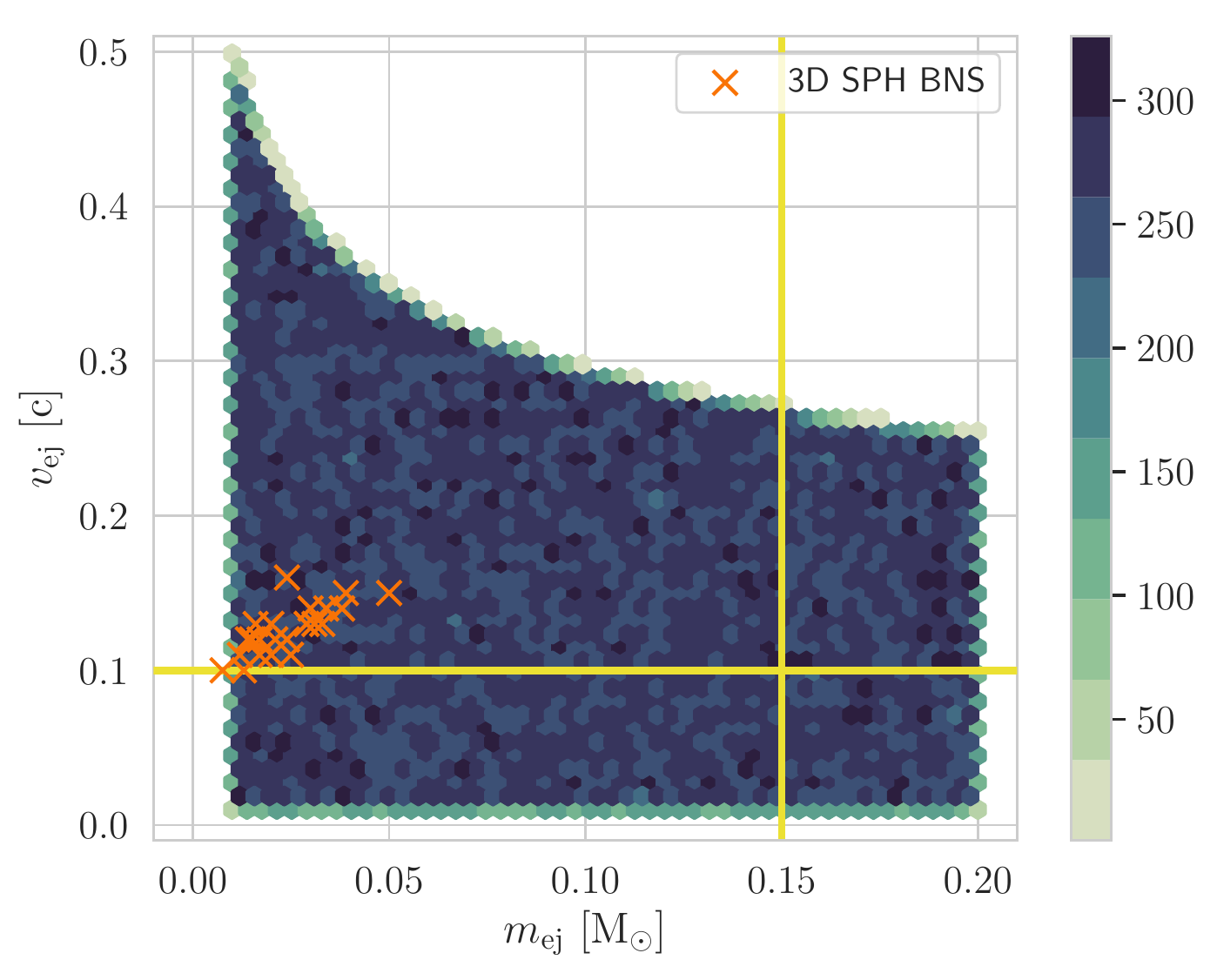}
  \caption{The sampling density of the two continuously varying parameters that characterize the population of BNS kNe from the SAEE (Population BNS) model. Points in this plane, representing individual kNe, are binned to show that for a given realisation of the kN distribution this space is uniformly sampled within the boundaries. The intersection of the two lines indicates the location of event GW170817 within the parameter space for a lower-limit `good fit' to the bolometric light curves, $({v_\mathrm{ej}} = 0.1 \, \mathrm{c}\,;\, {m_\mathrm{ej}} = 0.15\, \mathrm{M_{\odot}})$ as discussed by \protect\citet{Rosswog2018}. Overlaid are the locations of resulting ejecta parameters from the 3D smoothed-particle hydrodynamics (SPH) simulations of \protect\citet{Rosswog2012}. This indicates the range of parameters that BNS mergers were expected to exhibit as informed by simulations prior to GW170817.}\label{fig: ross_params}
\end{figure}

This model uses three parameters: the grey opacity ($\kappa$); the median ejecta mass (${m_\mathrm{ej}}$); and the median ejecta velocity (${v_\mathrm{ej}}$) to generate the time-series SED evolution for a kN event. The range of these parameters is shown in \autoref{tab: ross_params}. Without an observed population of BNS or NSBH mergers to inform our parameter selection, we adopt the generous parameter ranges considered by \citet{Rosswog2016a}, noting that this extends to the most optimistic cases seen by \citet{Rosswog2005} and \citet{Foucart2014} in simulations. \autoref{fig: ross_params} shows the sampling density of this range for the BNS case and illustrates where GW170817 would approximately fit within the parameter space. The upper bound on the ejecta velocity is motivated by ejecta kinetic energies of less than $10^{52}\, \mathrm{erg}$ that are typically found in simulations, e.g., table 2 of \citet{Rosswog2012}. This was also the approximate upper bound adopted in the exploration of the ${v_\mathrm{ej}}\,{-}\,{m_\mathrm{ej}}$ parameter space with numerical hydrodynamics simulations; see fig.\ 4 of \citet{Rosswog2016a}.

The distribution of grey opacity is linked to the three-dimensional geometry of the merger and the viewing angle of the observer in relation to the merger plane of the event. Not only are there modelling uncertainties for the geometry and the mapping between viewing angle and opacity, but there is a high degree of uncertainty in the opacities that are determined by the line transitions of complex ions \citep{Kasen2013}. In particular, lanthanide ions can have enormous numbers of lines (greater than $10^7$) whose energy levels are not accurately known. For ejecta mixtures containing a substantial fraction of lanthanides (produced by ejecta with electron fractions below $Y_e =0.25$), we adopt high opacities of $\kappa = 10\,\mathrm{cm^2 \, g^{-1}}$ \citep{Kasen2013, Tanaka2013}. For lower lanthanide-containing ejecta compositions, i.e., greater values of $Y_e$, we use $\kappa = 1\,\mathrm{cm^2 \, g^{-1}}$ \citep{Rosswog2016a}. For these reasons, we adopt a binomial opacity distribution. With our isotropic one-dimensional SAEE model this effectively maps onto viewing the event face-on or edge-on, i.e., perpendicular to or aligned with the merger plane, for the low and high opacity cases respectively.

For the NSBH case we also use the SAEE model but consider slightly different parameter ranges. These events are expected to be more energetic in their dynamical mass loss but lack a low opacity blue component \citep{Rosswog2016a, Metzger2016}. Tidal forces are the dominant ejection mechanism of matter in NSBH mergers (rather than neutrino-driven winds) and therefore the ejecta are expected to have the very low electron fraction of the original neutron star material. Thus, we only consider the high opacity ($\kappa = 10\,\mathrm{cm^2 \, g^{-1}}$) case with increased lower limits on the median ejecta mass and median ejecta velocity. The relevant parameter ranges for NSBH kNe are also shown in \autoref{tab: ross_params}.

\begin{figure}
  \centering
  \includegraphics[width=\columnwidth]{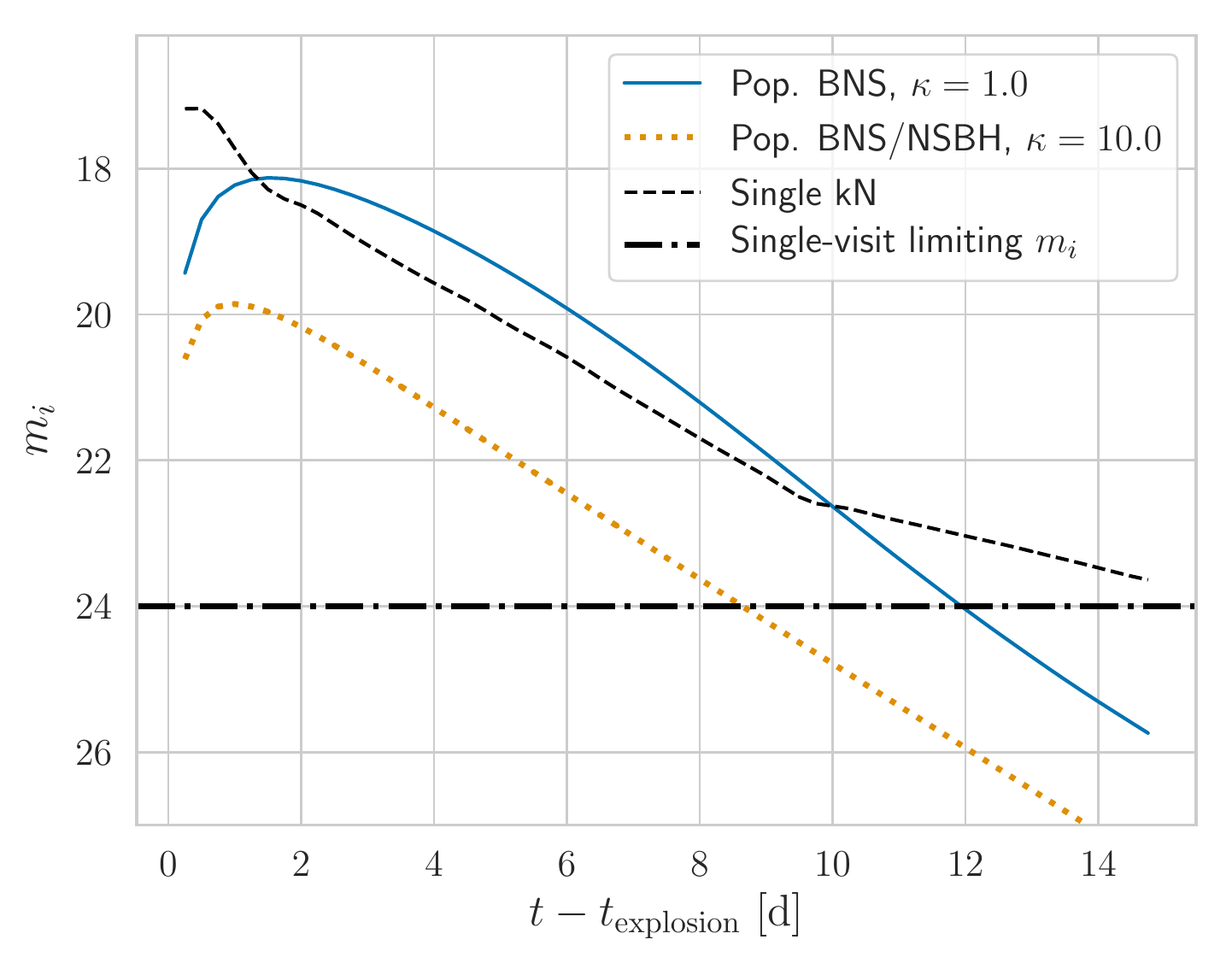}
  \caption{Model light curves, in the LSST $i$-filter at a redshift of $z=0.01$, for all three kN models considered in this work. The ejecta parameters that generate the light curves from the Population BNS/NSBH model are set to those of GW170817, see Fig.\ \ref{fig: ross_params}. For identical ejecta parameters, the lower opacity Population BNS case is much brighter than the higher opacity case, but neither model includes the early `bump' from GW170817 \protect\citep{Villar2017b}. In general, as the Single kN model does not vary with a choice of parameters, the Single kN model is brighter than the light curves from both Population BNS/NSBH models for most of the parameter space considered.}\label{fig:model_compare}
\end{figure}

To distinguish between the three kN models described above, throughout the rest of this work we will refer to the kN model based on the DES observations of GW170817 as the `Single kN' model and the SAEE models for BNS and NSBH kNe as the `Population BNS' and `Population NSBH' models respectively. These represent two scenarios. One is an optimistic scenario, where the high-luminosity Single kN model, with both blue and red-kN components, represents all events. The second is a more conservative and physically motivated scenario, where either of the Population BNS/NSBH models describe a population of kN events, of which the majority are less luminous than the Single kN events. Fig.\ \ref{fig:model_compare} shows LSST $i$-filter light curves for the Single kN model and the Population BNS/NSBH models, with ejecta parameters of the population models equal to the lower-limit `good fit' to the bolometric light curves of GW170817 discussed by \citet{Rosswog2018}, $({v_\mathrm{ej}} = 0.1\,\mathrm{c}\,;\, {m_\mathrm{ej}} = 0.15\, \mathrm{M_{\odot}})$.

\vspace{-0.5\baselineskip}

\subsection{Transient Distribution}\label{subsec:cosmo_dist}
To simulate observations of kNe, we begin with the number of events $N_{\mathrm{total}}$, per comoving volume $V_{\mathrm{co}}$, per event rest-frame time, $t_{\mathrm{rest}}$; i.e., the comoving event rate density
\begin{equation}\label{eqn:erd}
   \Gamma_{\mathrm{co}} =\frac{\md N_{\mathrm{total}}}{\md V_{\mathrm{co}}\, \md t_{\mathrm{rest}}}\,.
\end{equation}
This rate, $\Gamma_{\mathrm{co}}$, is assumed to be constant over the late-time cosmic evolution we consider. For BNS mergers we assume a rate of $1000 \, \mathrm{Gpc^{-3} \,yr^{-1}}$ \citep{Scolnic2017a}; this rate falls within the most recent rate estimates for BNS mergers from LIGO/Virgo, which place the 90 per cent confidence interval on these rates at $110{-}3840\, \mathrm{Gpc^{-3} \,yr^{-1}}$ {\citep{TheLIGOScientificCollaboration2018}}. This rate is also consistent with the rate derived by \citet{DellaValle2018b} using the rates of BNS mergers after GW170817, short-GRBs, and upper-limits from optical observations. The NSBH merger rate we choose is $300 \, \mathrm{Gpc^{-3} \,yr^{-1}}$. This is based on our choice of the BNS rate scaled by the ratio of the rate upper limits from \citet{TheLIGOScientificCollaboration2016}. This is also within the upper-limit rate of the most recent LIGO/Virgo observing runs, which currently place a 90 per cent confidence upper-limit at $610 \, \mathrm{Gpc^{-3} \,yr^{-1}}$ \citep{TheLIGOScientificCollaboration2018}.

To compute the redshift distribution of events in the observer's frame, the distribution of events in redshift space $n_{\mathrm{events}}$ and the cumulative number of events $N_{\mathrm{total}}$, observed out to some redshift, are defined. These two quantities are related by
\begin{equation}\label{eqn:total_dist}
N_{\mathrm{total}}(z) = \int_0^{z} n_{\mathrm{events}}(z') \, \md z'\,.
\end{equation}
Further, using Eq. \ref{eqn:erd} we obtain
\begin{equation}
N_{\mathrm{total}}(z) = \int_0^{T (z)} \int_0^{V_{\mathrm{co}}(z)}  \Gamma_{\mathrm{co}} \, \md V'_{\mathrm{co}}(z) \,\md t_{\mathrm{rest}} \, ,
\end{equation}
where $T(z)$ is the total time elapsed in the rest frame at redshift $z$. In terms of quantities directly connected to properties of the observations (observation time-interval, $\Delta T_{\mathrm{obs}}$, redshift, $z$, and the sky area, $\Delta \Omega_{\mathrm{obs}}$) the redshift distribution is
\begin{align}\label{eqn:reddist}
n_{\mathrm{events}}(z) = \, c^3  \frac{\Gamma_{\mathrm{co}}}{(1+z)H(z)} \left[\int^{z}_0 \frac{\md z'}{H(z')} \right]^2 \, \Delta T_{\mathrm{obs}} \, \Delta \Omega_{\mathrm{obs}}\,.
\end{align}

The number of observed events then follows a Poisson distribution, for which the cumulative distribution function is
\begin{equation}\label{eqn:invcum}
P(k;N_{\mathrm{total}}(z)) =e^{-N_{\mathrm{ total}}(z)} \sum_{i=0}^k \frac{N_{\mathrm{ total}}(z)^i}{i!}\,,
\end{equation}
and $k$ is the recorded count. Now, choosing a maximum redshift $z_{\mathrm{max}}$ and computing $N_{\mathrm{ total}}$, we obtain a realisation of the observed events. As this is a discrete distribution, an approximation to the inverse of the cumulative probability distribution, Eq. \ref{eqn:invcum}, must be used. This is done by randomly drawing from a uniform unit interval distribution, $u\in(0,1]$, which represents evaluations of the cumulative distribution function. Then the value of $k$ is found that gives the closest value to the draw, $u$. This $k$ corresponds to a realisation of the total number of events from the Poisson distribution, $N_{\mathrm{ total}}^{\mathrm{realisation}}$. With this draw of the total number of events, it is now possible to build a realisation of the redshift distribution.

Given the total number of events $N_{\mathrm{total}}(z_{\mathrm{max}})$, the cumulative redshift distribution of events $N_{\mathrm{total}}(z)$ is normalized to the total number of kNe. This yields a cumulative probability distribution function of kNe as a function of redshift, $F(z) = N_{\mathrm{total}}(z)/ N_{\mathrm{total}}(z_{\mathrm{max}})$. Next, numbers are drawn from a uniform unit interval distribution, $u_i\in(0,1]$, where $i$ runs from one to $N_{\mathrm{ total}}^{\mathrm{realisation}}$. Then, using the inverse of the cumulative distribution function, these draws are mapped from the uniform distribution to the redshifts that generate these values, $F^{-1}(u_i) \rightarrow z_i$. This produces a set of redshifts for all the kNe that occur within the observer frame represented by $\Delta \Omega_{\mathrm{obs}}$, $\Delta T_{\mathrm{obs}}$, and the redshift range, $z \in [0,z_{\mathrm{max}}]$. For our work we set $z_\mathrm{max} = 0.75$. This is the redshift where the brightest kN we simulate is below an SNR of one in all LSST filters given the limiting magnitudes that are expected \citep{TheLSSTDarkEnergyScienceCollaboration2018}.

With the number of events and their redshift distribution calculated, we place the kNe randomly on the sphere. With a prior that is uniform per solid angle, the events are placed in right ascension and declination within the declination band that covers the entire LSST-observed sky area for a given survey strategy. Additionally, the time of explosion for each event is chosen with a uniform prior over the survey lifetime; this incorporates an extension at the start of the survey to include events that would overlap with the survey at the end of their evolution. After these steps, the cosmological distribution of kN is fixed.

\vspace{-0.5\baselineskip}

\subsection{Gravitational Wave Signals}\label{subsec:GWsignals}
We also compute an approximate gravitational wave merger signal for the BNS sources in our distribution using the {\sc PyCBC} software \citep{DalCanton2014, Usman2016, Nitz2018}.\footnote{\url{https://pycbc.org/}} In order to map kN parameters to the relevant merger parameters we use those from the merger signal of GW170817.\footnote{Such a mapping does not yet exist for the Population BNS model and is the subject of future work.} We assume the same masses and spins for the simulated GW signals associated with both BNS kNe models. We obtain the relevant parameter values using the {\sc Merger} package of the {\sc PyCBC} software to access the median one-dimensional marginalised parameters from the low spin prior of \citet{TheLIGOScientificCollaboration2017}. This gives equal masses of $1.36 \, \mathrm{M_{\odot}}$ for both neutron stars and assumes that both objects are irrotational \citep{TheLIGOScientificCollaboration2017}. As no observed event is known for the NSBH case, and the large uncertainty in choosing what could be `typical' merger parameters, we do not compute NSBH gravitational waveforms. For the BNS mergers, we use parameters from the kN distribution and EM signal modelling: right ascension, declination, explosion time in observer frame as the merger time, luminosity distance, and inclination. The detector noise is assumed to be Gaussian and estimated using the design-sensitivity power spectral density of each detector \citep{Abbott2018}. This comprises the simulated detector signal on which we will perform matched filtering to compute the SNR.

\section{Simulations}\label{sec: sims}
\subsection{Survey Strategy}\label{subsec:survey_strat}
The LSST is being designed to perform four broad science cases: catalogue the Solar System, map the Milky Way, explore the time-domain sky, and place unprecedented constraints on dark matter and dark energy \citep{Ivezic2008,LSSTScienceCollaboration2009a}. To meet these goals the fiducial LSST survey is segmented into a main survey, the WFD, which comprises approximately 90 per cent of operations and the remainder of the time given to a set of mini-surveys to address the science cases not covered by the WFD \citep{LSSTScienceCollaboration2017}. The LSST Project is currently (2018) performing an exercise to determine an optimised survey strategy to maximise the return for all science cases.\footnote{\url{http://ls.st/doc-28382}} As these science cases are quite varied, nearly every aspect of the survey strategy is up for reconsideration. To test the impact of proposed survey changes, the LSST Project has provided a sizeable list of simulated alternative survey strategies\footnote{\url{http://astro-lsst-01.astro.washington.edu:8080/}} and provided tools to explore further changes not already represented in those provided.

The simulated survey strategies are generated with the LSST Project's Operations Simulator ({\sc OpSim}) software\footnote{\url{https://www.lsst.org/scientists/simulations/opsim}} \citep{Delgado2016, Reuter2016}, the Alternative Scheduler ({\sc Alt\_Sched})\footnote{\url{http://altsched.rothchild.me:8080/}}, and the Feature-Based Scheduler ({\sf FBS}) \citep{Naghib2018} also supplied by the LSST Project. These survey simulators generate a library of telescope pointings and a wealth of meta-information. Additionally, we use the dithered pointing locations for each {\sc OpSim} simulated survey generated with the proposed dithering strategy of \citet{Awan2016}.\footnote{Randomly shifting the telescope-pointing by a small amount, and randomly rotating the sensor plane by a small amount, ensures greater survey uniformity of depth across the sky \citep{Awan2016}.} The other survey simulators do not use a fixed grid to compute observation pointings, or already include dithering, and thus do not require translational dithering to be added \citep{Naghib2018}. The cadence library and, if applicable, the associated dithered pointing file are taken as inputs to our simulation package. Using the {\sc OpSimSummary}\footnote{\url{https://github.com/rbiswas4/OpSimSummary}} software, a summary table from each survey strategy database is obtained containing the dithered telescope-pointing locations, time of observation, sky noise, filter, and other information (Biswas et al. {\it in prep}). Collecting this information allows these simulated telescope pointings to be applied as a selection function on our kN distribution and simulate the instrument observations.

The survey strategies we have chosen, shown in \autoref{tab: cadence_names}, explore the major changes to the WFD survey strategy represented by the simulated surveys provided in the LSST Call for White Papers \citep{LSSTwpcall2018}. These proposed changes can be summarised as the following: larger footprint area, higher nightly cadence or `rolling'-style strategy,\footnote{A `rolling' survey strategy refers to splitting the main survey area into multiple declination regions and alternating which region is being actively observed \citep{LSSTScienceCollaboration2017}.} nightly re-observation of all locations in a different filter, alternative scanning strategy, and change in exposure time.  Multiple simulated survey strategies have been provided for each of these changes. In addition to the baseline strategy, we consider six other survey strategies that each capture one or more of these proposed changes. These survey strategies are also compatible with the recommendations put forth by the LSST Dark Energy Science Collaboration (DESC) \citep{Lochner2018a, Scolnic2018}.

\subsection{Observations}\label{sec: obs}
 For both the Single kN and Population kNe models, we simulate the observations as follows. As inputs we take the survey strategy summary table, information about the instrument including filter throughputs, per-filter flux zero-points, field-of-view geometry, and maximum redshift. The cosmological distribution of kNe is generated as explained in Sec. \ref{subsec:cosmo_dist}, for which we use the cosmological parameters from \citet{PlanckCollaboration2015a}. Then, given a choice of kN model, we generate a time-series SED for each event. Depending on the combination of redshift, right ascension, and declination that it has been assigned, the SED is redshifted, dimmed, and modified with dust extinction according to the extinction map $E(B{-}V)$ using LSST per-filter corrections from \citet{Schlafly2011}. We obtain $E(B{-}V)$ values using the software package {\sc sfdmap}.\footnote{\url{https://github.com/kbarbary/sfdmap}} Each object is also assigned a peculiar velocity that Doppler-shifts the SED in the event rest frame, such that $1 + z_{\mathrm{obs}} = (1+z_{\mathrm{cosmo}})(1+z_{\mathrm{pec}})$. A summary of these parameter choices is presented in \autoref{tab: sim_params}.

\begin{table}
\centering
\resizebox{\columnwidth}{!}{%
  \begin{tabular}{c|c|c}
  Parameter & Values & Note \\
  \hline
$z$ & $[0, 0.75]$ & Set by SNR $\leq1$ of brightest kN. \\
 \hline
$R_{\mathrm{BNS}}$ & $1000 \, \mathrm{Gpc^{-3} \,yr^{-1}}$ & Rate used by \citet{Scolnic2017a}. \\
\hline
$R_{\mathrm{NSBH}}$ & $300 \, \mathrm{Gpc^{-3} \,yr^{-1}}$ & See Sec.\ \ref{subsec:cosmo_dist} \\
\hline
Dust Map & Per-filter extinction & \citet{Schlafly2011} \\
\hline
Peculiar Velocity & N($\mu = 0; \sigma = 300 \, \mathrm{km\, s^{-1}}$) & `Typical' values \citep{Davis2010}.\\
\hline
  \end{tabular}
}
 \caption{Parameters used to generate the simulated LSST kNe observations.}\label{tab: sim_params}
\end{table}

The choice of these parameters and the computed kN distribution fix the EM signals simulated for the entire survey duration. With this distribution of light curves, the chosen survey strategy samples these signals to create mock observations of each event. For a given kN, all telescope-pointings that overlap with the event's spatial location and lifetime are found. Using telescope-pointing information, such as the filter five-sigma limiting magnitude, and assuming a circular field-of-view geometry the instrument-measured flux, uncertainties, and other properties of the observations are computed. We additionally compute the nightly coadded observations per filter and use this as an additional dataset. Both sets of simulated observations are then processed through the sets of detection criteria discussed in Sec.\ \ref{sec: criteria} and App.\ \ref{sec:appendix}.

The subset of kNe that pass all conditions for a given set of detection criteria are labelled as detections. The results we report use a single realisation of the kNe redshift distribution for each survey strategy. Each of these realisations represents a draw from a Poisson distribution of kN events that has been sampled with a survey strategy. The major contribution to the uncertainty in the number of reported detections, given a comoving event rate density, is the Poisson uncertainty. For our results this is well-approximated by the square-root of the number of detections. To cross-check our process, we ran a simulation with our software using the configuration used by \citet{Scolnic2017a} with the observation simulation software {\sc SNANA} \citep{Kessler2009}. We found that our results agreed within the uncertainty, yielding 67 kNe over the ten year survey in comparison to the 69 kNe reported by \citet{Scolnic2017a}.

\subsection{Gravitational Wave Signal-to-Noise Ratio}
Lastly, for the set of kN which pass the sets of detection criteria discussed below, we compute the gravitational waveforms outlined in Sec. \ref{subsec:GWsignals} and perform matched filtering to compute the SNRs in the Advanced LIGO Livingston and Hanford observatories, the Advanced Virgo detector, and their combined, ALV, network SNR. To do this, we use the software {\sc PyCBC} \citep{Nitz2018}. As we are only considering BNS inspirals without component spins or tidal deformations using a post-Newtonian approximation is valid \citep{Blanchet2013, Cho2018}. The waveform approximant we use is {\sf TaylorF2}. It is analytic and computationally inexpensive, making it a commonly used waveform for designing template banks to search GW data for BNS inspirals \citep{Canton2017}. For each detector we assume a noise level corresponding to the detector's design sensitivity and add the noise into the merger waveform. We then perform matched filtering to compute a time-series of SNRs for each detector. We take the maximum of these SNRs for each individual detector and calculate the network SNR as the square root of the sum of the squares of the individual SNRs. To determine if a GW signal is detected by the ALV network, we adopt the commonly used threshold for GW detection of an $\mathrm{SNR}$ of 12 \citep{Huang2018}.

\section{Detection Criteria for the Electromagnetic Counterpart}\label{sec: criteria}
There does not exist an accepted set of criteria for kNe for which a detection could be claimed if all criteria are passed. With only one known kN, it has not yet been possible to test different detection criteria on real data. Detection criteria allow us to estimate the number of kNe for which we will have sufficient information to identify them and provide a trigger to GW observatories to search for sub-threshold events. The set of detection criteria we use is that of \citet{Scolnic2017a}:
\begin{itemize}
  \item Two alerts separated by $\geq 30$ minutes.
  \item Observations in at least two filters with $\mathrm{SNR} \geq 5$.
  \item Observations with $\mathrm{SNR} \geq 5$ separated by a maximum of 25 days.
  \item A minimum of one observation of the same location within 20 days before the first $\mathrm{SNR} \geq 5$ observation.
  \item A minimum of one observation of the same location within 20 days after the last $\mathrm{SNR} \geq 5$ observation.
\end{itemize}

These address several issues that arise when determining a detection. The first criterion is intended to reject asteroids and identify true astrophysical transients \citep{Scolnic2017a}. The second criterion sets a minimum on the quality of light curve information any object must have to be detected, which can be used to discriminate between transient types. The third criterion rules out bright, extended-duration light curves like superluminous supernovae or active galactic nuclei \citep{Scolnic2017a}. The final two criteria effectively require a minimum of four light curve points per object and ensure the event's maximum is within the active observing season \citep{Scolnic2017a}.

In the context of these detection criteria, an alert is an observation that has a measured SNR effectively greater than five after difference-image subtraction \citep{Kessler2015}. We do not simulate images or template subtraction. Thus, our simulated SNRs do not correspond to those found through this process. To mimic alert generation from template subtraction, we make simulations using the per-filter alert efficiency vs.\ true SNR ratio response function. True SNR refers the SNR measured from the catalogue source-flux over the uncertainty of the sky noise. As these per-filter response functions are unknown for LSST prior to operation, we have used the response functions that were adopted for the DES Y1 analysis \citep{Kessler2015}. Given the DES filter-set is similar to that of LSST, and the variation of the response function between filters is small, this is not an unreasonable approximation.

\begin{figure}
  \centering
  \includegraphics[width=\columnwidth]{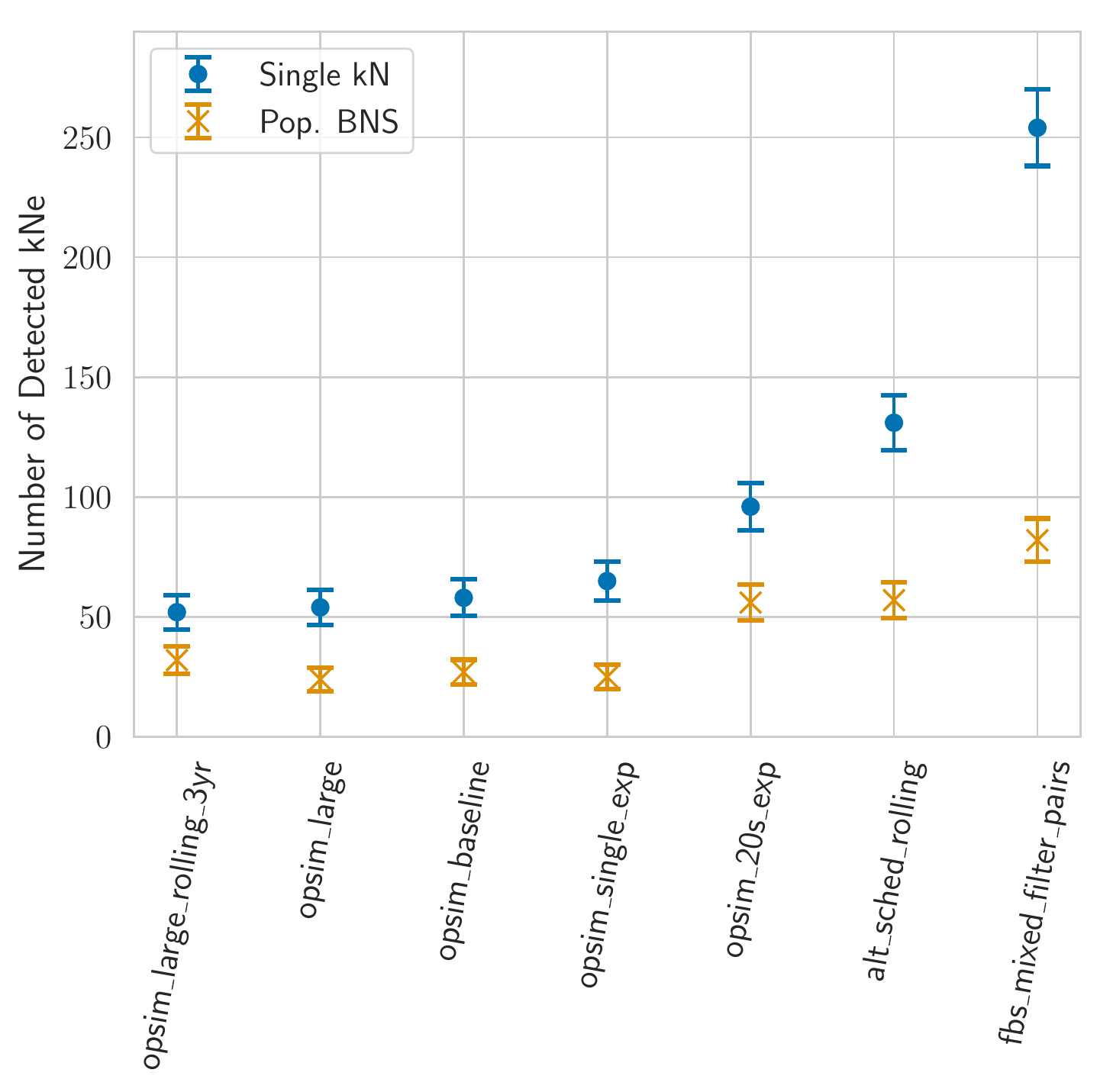}
  \caption{Comparison of the ten-year detection counts, according to the detection criteria S2, for each survey strategy and both BNS kN models. The predicted number of detections is greater for the Single kN model, but both models show generally the same variation with the choice of survey strategy. Both share a large peak in the number of detected kNe for the survey strategy {\sf fbs\_mixed\_filter\_pairs}, and two further survey strategies show marked improvement over the baseline strategy.}
  \label{fig:BNS_compare}
\end{figure}

\begin{figure}
  \centering
  \includegraphics[width=\columnwidth]{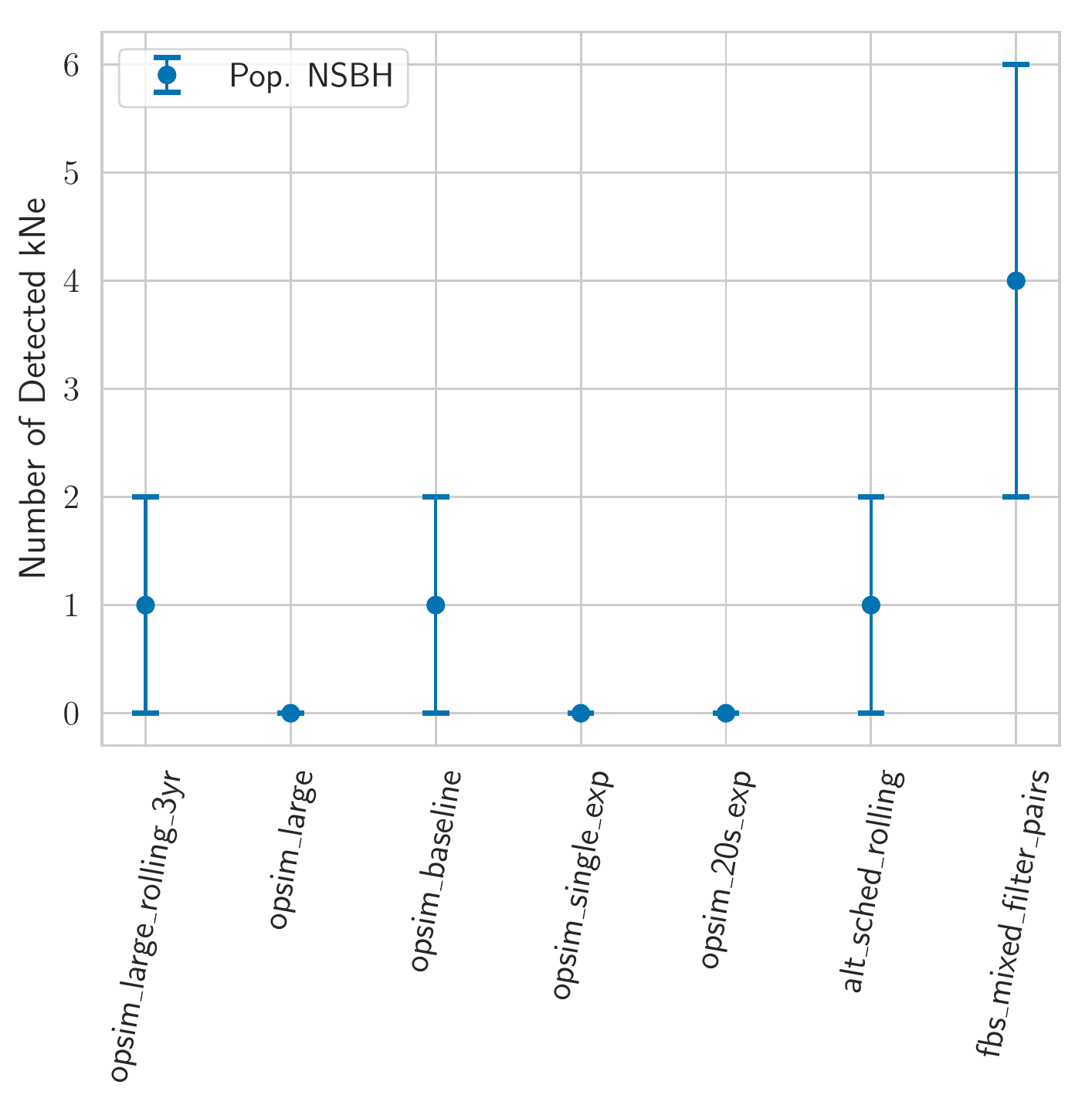}
  \caption{Ten-year detection counts, according to the detection criteria S2, for the Population NSBH kNe model for each survey strategy. The detection counts for NSBH mergers are very low. We expect detections to be lower due to their reduced rate and dimmer EM signature.}
  \label{fig:NSBH}
\end{figure}

\begin{figure*}
  \hspace*{-0.9cm}
  \centering
  \includegraphics[width=0.8\textwidth]{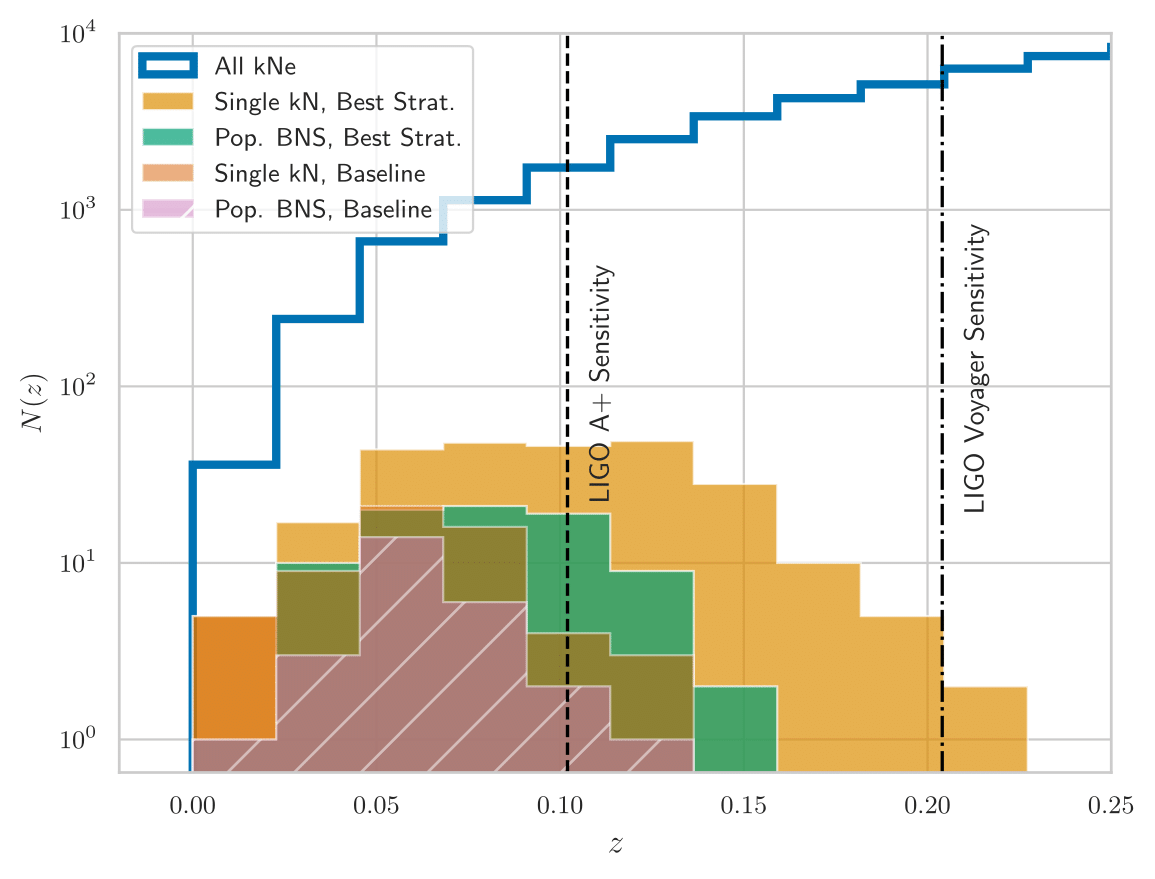}
  \caption{Example redshift distributions of serendipitously detected kNe, given the detection criteria S2, after the full ten-year survey. This histogram compares the baseline strategy, {\sf opsim\_baseline}, with the best strategy, {\sf fbs\_mixed\_filter\_pairs}, for both the Single kN and the Population BNS kN models. The redshift distribution of all simulated kNe is also shown. The design sensitivities of future GW observatories, given the most optimistically configured BNS merger scenarios, are shown by vertical lines \protect\citep{Chen2017a, Chamberlain2017}. The best LSST survey strategy significantly increases the range of the redshift distribution for both kN models. As shown in App.\ \ref{sec:appendix}, some kNe are detected by LSST below the detection threshold of ALV for all survey strategies. However, if an optimal survey strategy is adopted it will be possible to detect kNe beyond the best-case sensitivity of future detectors that will be operational during the time of LSST, such as LIGO A+.}
  \label{fig:typical_nz}
\end{figure*}

\citet{Scolnic2017a} considered this set of detection criteria, using observations that were coadded nightly. However, it is not expected that nightly coadded observations will be made available by the LSST Project for the WFD survey; see the Data Products Definition Document \citep{LSSTDDP2018}. While it is possible to do the nightly coadds, this will take additional computational resources that, at this time, will not be part of the LSST data management pipeline \citep{LSSTDDP2018}. For this reason, we prefer detection criteria evaluated on individual observations, not coadds, though we report results from both. We label detection criteria in the following way: the criteria of \citet{Scolnic2017a} evaluated on coadded observations is referred to as S1 and the criteria of \citet{Scolnic2017a} evaluated on individual observations is S2. The results from all sets of criteria, including the detection criteria of \citet{Cowperthwaite2018}, are tabulated in App.\ \ref{sec:appendix} for comparison with other works.

\section{Results}\label{sec: results}
\begin{figure}
  \centering
  \includegraphics[width=\columnwidth]{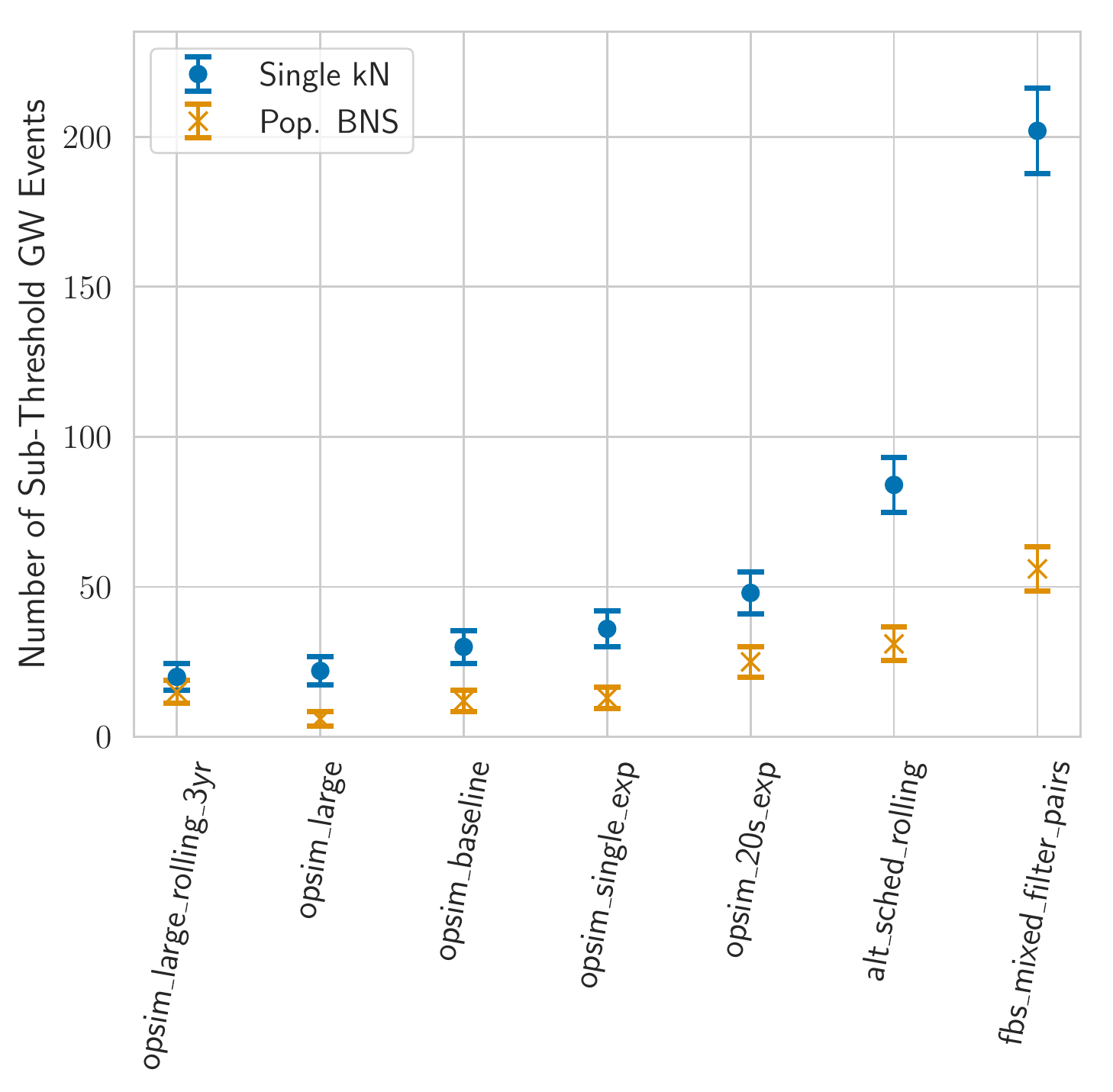}
  \caption{The number of kNe detected with LSST, given the detection criteria S2 from Fig.\ \ref{fig:BNS_compare}, that are also below the GW signal detection threshold of $\mathrm{SNR}=12$ if observed by the ALV detector network operating at design sensitivity. Assuming continuous operation of the network conservatively yields an upper-limit on the number of kNe that could be used to trigger searches for sub-threshold events in GW data. Comparing to Fig.\ \ref{fig:BNS_compare}, we see that the numbers in this subset trace the overall detection counts, with the largest subset also corresponding to the highest detection counts from the survey strategy {\sf fbs\_mixed\_filter\_pairs}.}
  \label{fig:gw_counts}
\end{figure}

The number of detections for each kN model, according to the detection criteria S2, are shown in Fig.\ \ref{fig:BNS_compare} and Fig.\ \ref{fig:NSBH}. As discussed in Sec.\ \ref{sec: obs}, the uncertainty reported for all detection results is the Poisson uncertainty. For BNS kNe, the detected numbers are greater than baseline by a factor of more than 2 for three survey strategies and relatively flat, within 20 per cent with respect to the baseline strategy, for the other three survey strategies considered. For NSBH kNe we find low numbers of $0{-}4$ over the survey lifetime. The difference in redshift distributions of detected kNe, for the baseline and the optimal strategy, are shown in Fig. \ref{fig:typical_nz}. We see that from the baseline strategy to the optimal strategy this range increases from $0 \leq z \leq 0.15$ to $0 \leq z \leq 0.22$. These redshift ranges are representative of those found for other strategies. The lower redshift range corresponds to the strategies obtaining detection counts similar to the baseline strategy. The larger redshift range similarly corresponds to the survey strategies which see improved detection numbers over the baseline.

The effect of changing the kNe model can also be seen in Fig. \ref{fig:typical_nz}. It shows that at all redshifts, the detection numbers are suppressed for the Population BNS model in comparison to the Single kN model. Additionally, in App. \ref{sec:appendix} we tabulate detection counts assuming the case of nightly coadded observations (see Fig. \ref{fig:det_met_compare}) and according to the detection criteria used by \citet{Cowperthwaite2018}. For BNS kNe we also have the computed the ALV GW network SNR. The numbers of detected kNe according to S2 that are below the detection-threshold of the ALV network are shown in Fig. \ref{fig:gw_counts}. The tabulated detection numbers and sub-threshold GW detections, for all survey strategies, models, and detection criteria are presented in \autoref{tab:all_results1}, \autoref{tab:all_results2}, and \autoref{tab:all_results3}.

\section{Discussion}\label{sec: discussion}
\subsection{Variation with Survey Strategy}
We find that three surveys show a marked improvement over the baseline strategy for serendipitous detections of kNe. The largest increase in detection numbers occurs for the {\sf fbs\_mixed\_filter\_pairs} strategy, the second for the {\sf alt\_sched\_rolling} survey strategy, and the third for {\sf opsim\_20s\_exp}. The first two of these include the change with respect to the baseline strategy of using a different filter for the required repeat observation of each sky location within a given night. If a location is scheduled to be observed which overlaps with a source near peak flux, obtaining observations in different filters within the same night will increase the likelihood of getting a high SNR multi-band measurement and multiple high-quality light curve data points.

This suggests that the most prominent factor prohibiting detections of kNe is the lack of multi-band information. This increase is not unexpected, as this change directly satisfies criterion two in both S1 and S2, and highlights the importance of the choice of detection criteria. Interestingly, both of these strategies include more frequent observations in different filters, yet {\sf fbs\_mixed\_filter\_pairs} performs much better than {\sf alt\_sched\_rolling}. Their relative counts indicate that after obtaining multi-band observations, most kNe have sufficient light curve points to pass detection. Thus, a rolling-style cadence misses more kNe due to the decrease in actively-observed sky area, compared with the number of detections gained due to more frequent observations of any sky location.

\begin{figure}
  \centering
  \includegraphics[width=\columnwidth]{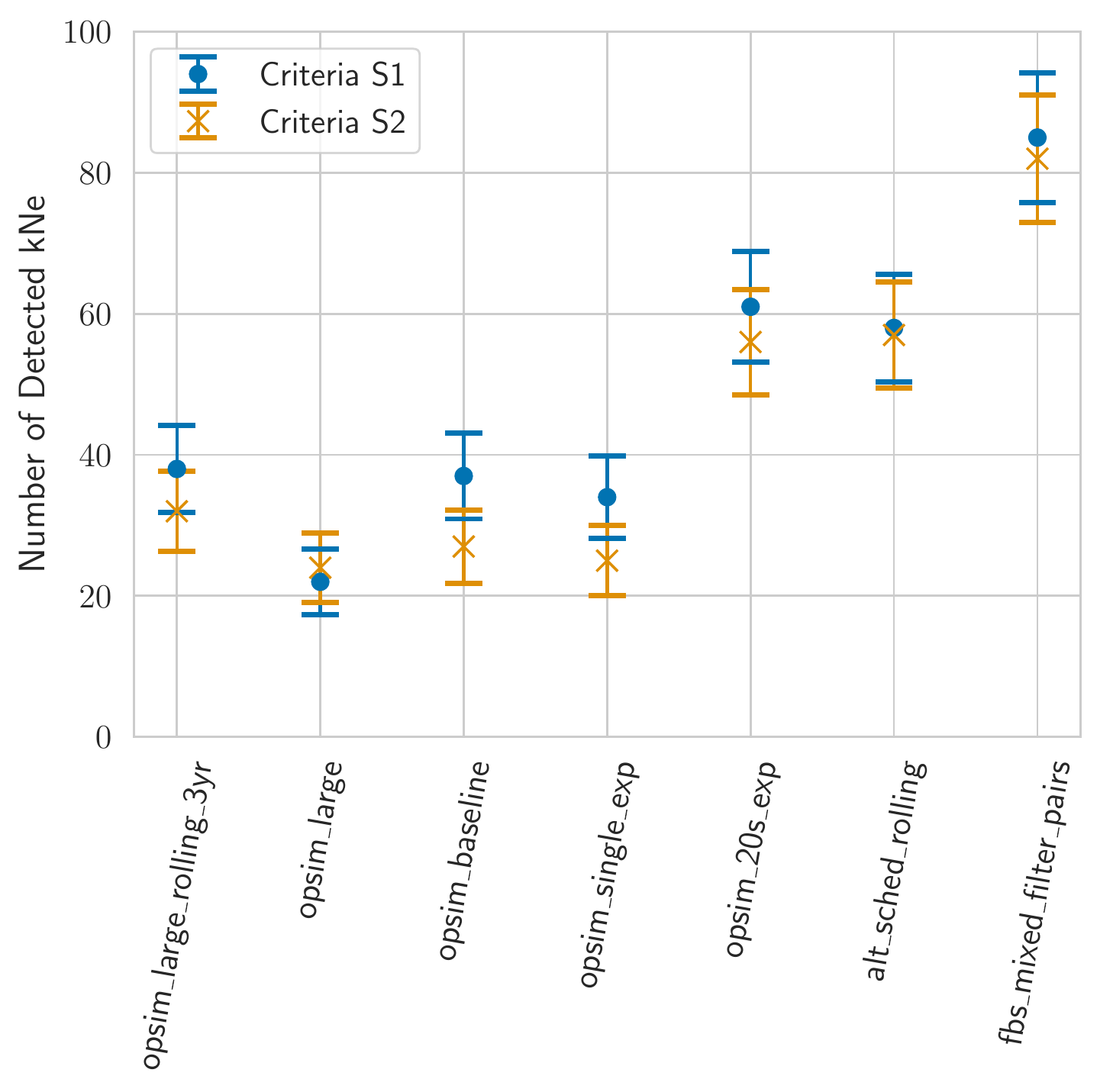}
  \caption{Comparison of ten-year detection counts per survey strategy for the Population BNS kN model given different choices of detection criteria. This compares the difference between using coadded observations and non-coadded observations. As expected, for most survey strategies, coadded observations yield more detections due to the increase in depth.}
  \label{fig:det_met_compare}
\end{figure}

The third strategy which offers improvement over the baseline, {\sf opsim\_20s\_exp}, explores the proposed change of reduced exposure time per observation, changing from a pair of back-to-back 15-second exposures in all filters to a single 20 second exposure in $griz$ and a single 40 second exposure in $uy$. As the median fraction of all observations in $griz$ for the baseline strategy is 73 per cent \citep{LSSTScienceCollaboration2017}, the reduction in time that the telescope spends taking an exposure at a single location increases the total number of observations each night more than the decrease that would be attributed to the longer $uy$ exposure times. This strategy maintains the nightly pair of observations in the same filter. Regardless, increasing the total number of observations while keeping the sky area being observed constant will decrease the average time between observations of a given sky location in all filters. Effectively this is again an increase in the cadence of obtaining observations in differing filters, like the other two strategies which improve the number of detections, though not to the same degree.

Given the finding that increasing the cadence of multi-band observations increases the number of kNe detections, we would also expect the survey strategy {\sf opsim\_single\_exp} to see an increased number of detections over baseline. This survey strategy removes the requirement of returning to a previously observed sky location within a single night. By removing the repeat observation, more of the observable sky will be covered in a single night. This reduces the gap in time between nights before which the telescope will be able to return to that location, potentially in a different filter. However, if we consider observations within a fixed time interval such as the duration of a kN light curve brighter than an SNR of five, removing the required pair of observations will increase the minimum number of nights needed to satisfy the detection criteria. For example, if a pair of observations in the same filter is required, in the best case scenario only two nights back-to-back are required to satisfy criteria one and two, see Sec.\ \ref{sec: criteria}. On one night the pair of observations is taken and on a second night the sky location is observed again, in a different filter. When a pair of observations is not taken, a minimum of three nights is needed to satisfy the detection criteria. As kNe evolve very quickly, increasing the number of nights needed for observations to satisfy the detection criteria decreases the number of detections. This negates the increase of detected kNe for this survey strategy due to the higher average cadence of obtaining multi-band information.

We now see that keeping the nightly pair of observations and increasing the cadence of multi-band observations improves the ability to detect kNe, by satisfying the most time-sensitive detection criteria, i.e., criteria one and two. If the pair of observations is taken using different filters, e.g., {\sf fbs\_mixed\_filter\_pairs}, this maximises the improvement that might be gained from a survey strategy; all of the observations that are relevant to satisfying the time-sensitive detection criteria are obtained in one night.

The other survey strategies considered do not implement changes which would lead to a significant modification to the cadence of obtaining observations in different filters. As such, the number of detected kNe for these strategies do not differ more than 20 per cent from the baseline survey strategy. Additionally, we find the dependence of BNS kNe detections on observing strategy is generally the same whether coadded or individual observations are used (see Fig. \ref{fig:det_met_compare}) though coadded observations generally detect more kNe due to the increase in depth.

\begin{figure}
  \centering
  \includegraphics[width=\columnwidth]{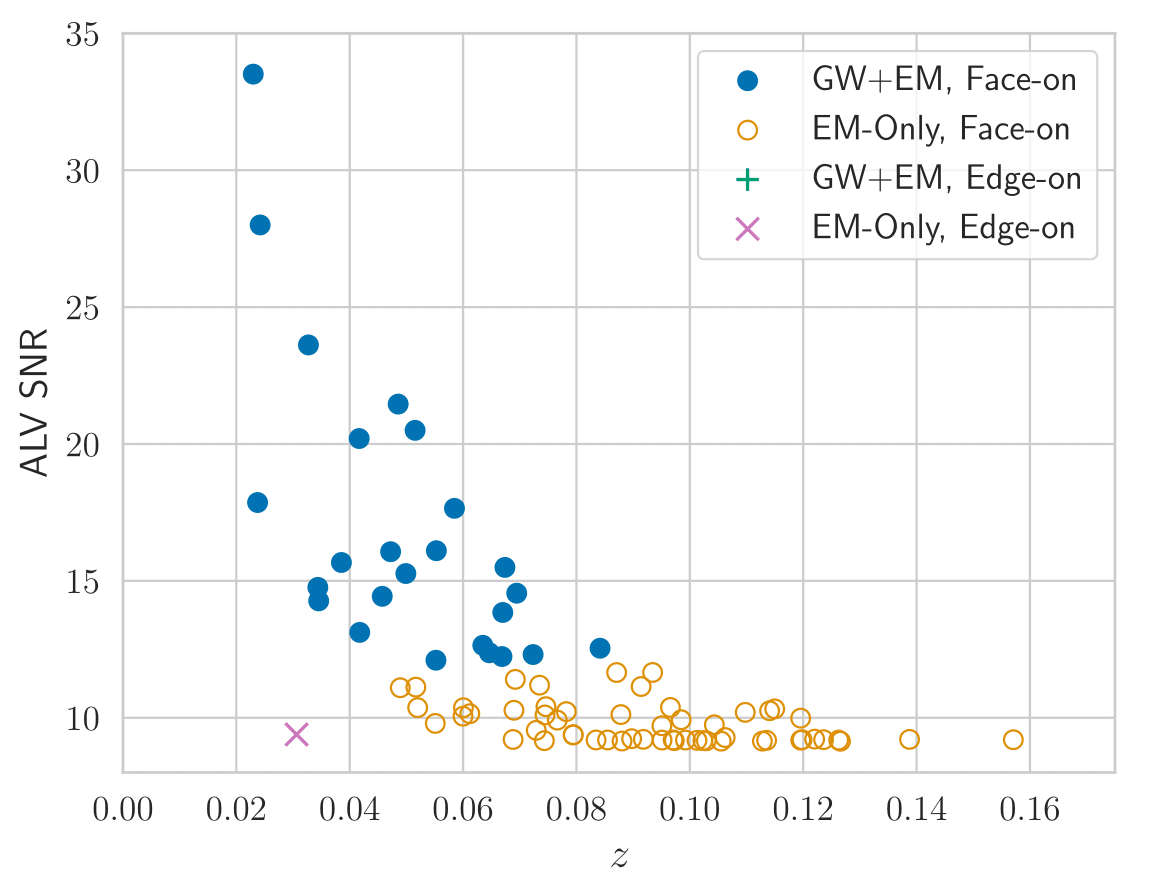}
  \caption{ALV SNR plotted against the source cosmological redshift for the sample of Population BNS kNe detected by LSST in the optimal survey strategy, {\sf fbs\_mixed\_filter\_pairs}, given the detection criteria S2. This sample is sub-divided into detections of only the kN component or detections of both the kN and the counterpart GW signal. Furthermore, this sample is sub-divided by whether the event is simulated as a face-on or edge-on BNS merger. We see that, even for this optimal choice of survey strategy, only one edge-on event is detected by LSST and this event is not detected by the ALV network.}
  \label{fig: selection}
\end{figure}

\subsection{Variation with Kilonova Model}
We also see a significant difference in the number of predicted detections between kN models. As expected for BNS kNe, the Single kN model predicts a greater number of detections than the Population BNS model. Looking back at Fig.\ \ref{fig: ross_params}, we see that a lower-limit `good fit' for the Single kN model would fall at the larger kinetic energy end of the ejecta parameter space, if ${E_\mathrm{kin}}\approx (1/2) \,{m_\mathrm{ej}} {v_\mathrm{ej}}^2$. Additionally, the Single kN SED from GW170817 includes both an early-blue component and a red-kN component \citep{Villar2017b}. Our Population BNS model does not include both components in a single simulated kN \citep{Rosswog2018}; it is composed of either a blue or red-kN corresponding to viewing the event face-on or edge-on, due to the binomial choice of opacity, see Fig.\ \ref{fig:model_compare} and Sec.\ \ref{sec:kne}. Given the Single kN model's approximate dynamical ejecta parameters within the Population BNS model parameter space, the Single kN model corresponds to a greater luminosity than a large part of the parameter space that generates the Population BNS SEDs. As GW170817 appeared to be brighter than most predictions from theory and simulations of BNS kNe, using a model based on this single event will lead to unrepresentative results. Thus, a physically motivated model of kNe should be used make observational predictions.

\begin{figure}
  \centering
  \includegraphics[width=\columnwidth]{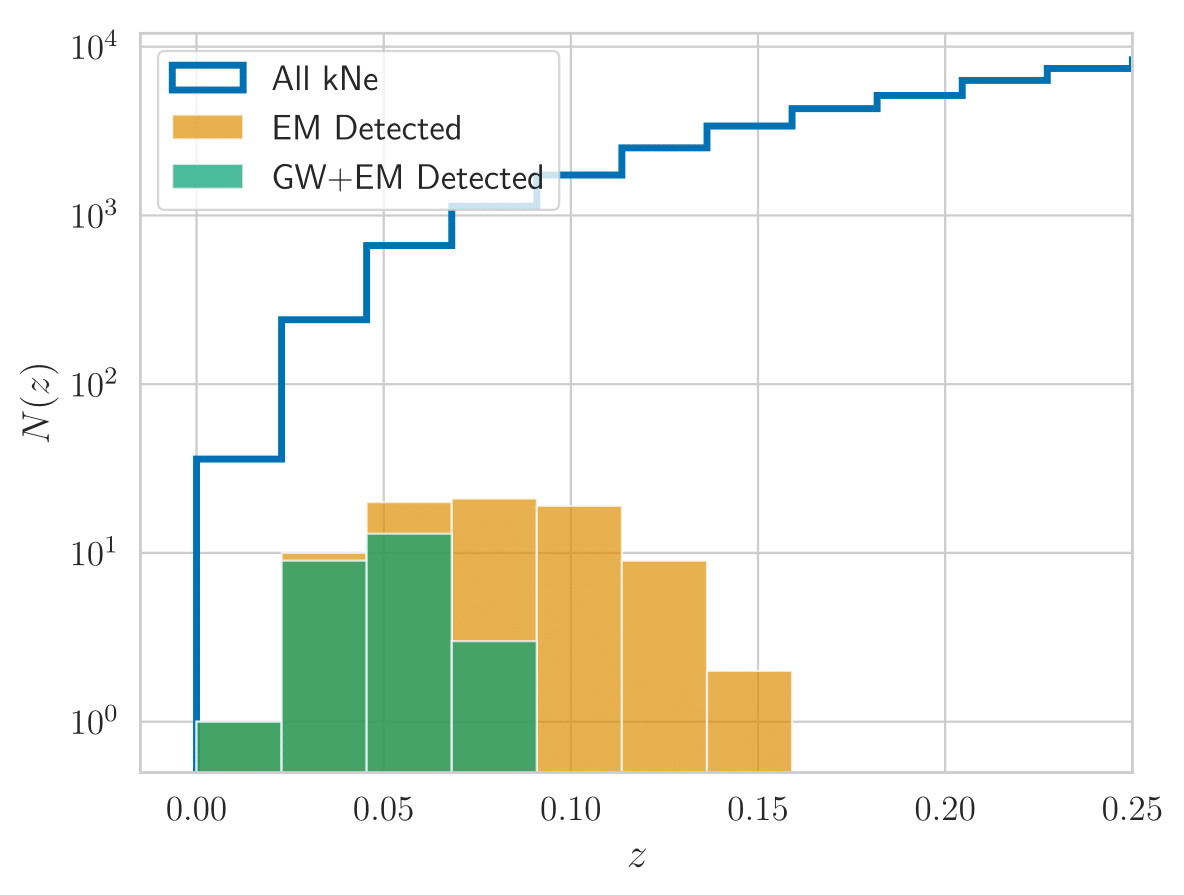}
  \caption{Example redshift distribution of Population BNS kNe that are detected after the full survey lifetime, given the detection criteria S2, for the optimal survey strategy, {\sf fbs\_mixed\_filter\_pairs}. The redshift histogram of the detected kNe is divided into two subsets. One subset, labelled GW+EM, are the LSST-detected kNe that are also above the GW SNR detection threshold if observed by ALV. The other subset, labelled EM-only, are the remaining kNe that are only detected by LSST and fall below the GW SNR threshold for detection. For this survey strategy, we see that there is a significant fraction, 68 per cent, of the kNe that do not pass this threshold.}
  \label{fig:typical_gw_nz}
\end{figure}

As shown in Fig.\ \ref{fig:NSBH}, even after the full ten year survey only a handful of NSBH kNe, at most, are detected; there appears to be no clear dependence on the choice of survey strategy. We do expect the number of NSBH detections to be fewer due to a lower merger rate for NSBH binaries. However, this large drop in the number of detections is worse than the 70 per cent reduction expected from the relative rate decrease. Considering again Fig.\ \ref{fig:model_compare}, for a given set of ejecta parameters NSBH kNe are much dimmer than a low opacity, i.e., face-on, BNS kNe. This points to the difficulty of detecting NSBH kNe and particularly the red-kN component. NSBH mergers are expected to only be accompanied by a red-kN and thus lack a blue, more luminous, component, as is expected for BNS \citep{Fernandez2016, Rosswog2016a}.

The modelling of the Population NSBH kNe is identical to the edge-on case of Population BNS model except for the higher lower bounds on ejecta parameters, see Table \ref{tab: ross_params}. If we consider again our assumption of either face-on or edge-on viewing of the BNS mergers, this suggests that events which occur face-on to our line-of-sight are preferentially detected. This then leads to a selection effect for face-on mergers of the population of kNe that are detected in both GW and EM waves, see Fig. \ref{fig: selection} for results from the Population BNS kNe model. This highlights the need to investigate further the detection of EM and GW signals for a range of viewing angles, but we leave this to future work.

\subsection{Sub-Threshold Gravitational Wave Detections}
For all survey strategies, the number of sub GW-threshold kNe detected by LSST, Fig.\ \ref{fig:gw_counts}, trace the overall counts of detected kNe shown in Fig.\ \ref{fig:BNS_compare} as we had anticipated. The increase in sub-threshold detections is due to the increase in redshift range of the detected kNe for survey strategies that perform better than the baseline, as shown in Fig.\ \ref{fig:typical_nz}. We see an increase of the redshift range corresponding to the optimal survey strategy {\sf fbs\_mixed\_filter\_pairs}, due to the cadence of multi-band observations. As previously mentioned, this survey strategy feature satisfies the detection criteria within a shorter period of time. Therefore, the detectability of kNe that spend less time above the limiting magnitude of the telescope, i.e., kNe at higher redshift, will be enhanced. To illustrate this point further, Fig.\ \ref{fig:typical_gw_nz} shows the redshift histograms for the Population BNS kNe detected according to S2 for the optimal strategy, and the subset that also are detected by the ALV network.

If we consider the results of all the GW detections, see Table \ref{tab:all_results1} and Table \ref{tab:all_results2}, all survey strategies will be able to detect some number of sub GW-threshold kNe. However, if we want to use these EM detections as triggers to search archival GW data, the number useable for this purpose will be smaller. This is due to the duty-cycles of GW detectors and the significant down-time for upgrades. We leave a more detailed analysis which accounts for the operational cycles of GW observatories during the LSST survey to future work.

\subsection{Contamination}
Another point worth exploring is the potential contamination from other transient types. \citet{Scolnic2017a} investigated this for contamination of the kNe sample by type Ia and core collapse supernovae and find only a three per cent contamination. More generally, \citet{Lochner2018a} showed survey strategies that perform better on classification and obtain larger numbers of `well-measured supernovae', also detect larger numbers of kNe. If a survey strategy is adopted that prioritises supernovae science and classification, this synergy will ensure low contamination of kNe detections from other transients.

\section{Conclusions}\label{sec: conclusion}
By simulating observations of BNS and NSBH kNe using realistic survey strategies, we showed the potential of LSST to serendipitously detect as many as $254 \pm 16$ kNe over the ten year survey if all kNe are GW170817-like or $82 \pm 9$ kNe assuming a representative population of kNe. For both models this is an improvement of more than 3 times the number detections that would be made with the current baseline survey strategy. The survey strategy which yielded this improvement was {\sf fbs\_mixed\_filter\_pairs}. This improvement occurred due to an increased cadence of obtaining multi-band observations of a given sky location.

The survey strategy feature which provided this improvement was the explicit requirement that the required repeat observation of a particular sky location within a night be done using a different filter than was used previously. The other survey strategy incorporating observation pairs in different filters within a single night also showed a significant improvement over baseline. However the numbers of detected kNe were lower in comparison to the optimal strategy due to the decreased sky area actively observed by a rolling cadence. Lastly, the survey strategy with a decrease of total exposure time per observation, while keeping other properties of the survey strategy fixed, also increased the cadence that a sky location is visited in any filter, though to a lesser degree. Correspondingly, this strategy saw a greater number of detections than the baseline strategy, but less than the optimal strategy. Ideally, a survey strategy designed for serendipitous detections would combine these two features to maximise the increased cadence of obtaining observations in different filters.

In addition to determining the number of serendipitous kNe detections, we also calculated the approximate SNR of the associated GW signals in the ALV GW detector network. The number of events which are detected by LSST and are sub-threshold for a GW detection directly corresponded to the survey strategies which also improved the total number of serendipitous detections. We found that, in the case of the optimal survey strategy, over the survey lifetime $202 \pm 14$ of the detected kNe for the optimistic GW170817-like model and $56 \pm 7$ kNe for the representative population model are sub-threshold GW events for the ALV detector network operating at design sensitivity. In either case this represents more than a 450 per cent increase over the baseline strategy.

The detection of such a population facilitates searches of archival GW data for these sub-threshold signals, increasing the available multi-messenger population. This will accelerate precision measurements of the Hubble constant, provide insight on the systematic uncertainties in such measurements, and explore the selection effects of GW detections of standard sirens. In the case that serendipitous detections of kNe are made during periods when no GW detectors are operational, these detections are also useful for accurately computing compact binary merger rates, and for studying the relation of the kN population to GRBs. With no other facility currently planned that will provide a comparable combination of speed, depth, and survey area, LSST will be a major facilitator of science with standard sirens.

\vspace{-0.5\baselineskip}

\section*{Acknowledgements}
This paper has undergone internal review in the LSST Dark Energy Science Collaboration. The internal reviewers were Renee Hlo\v{z}ek, Michelle Lochner, and Daniel Scolnic. We would like to thank Michelle Lochner, Daniel Scolnic, and the LSST DESC Observing Strategy Task Force for many useful discussions. Additionally, we thank Samaya Nissanke and Andrew Williamson for useful discussions regarding NSBH merger rates and the usage of the {\sc PyCBC} software. This research made use of {\sc Astropy},\footnote{\url{http://www.astropy.org}} a community-developed core Python package for Astronomy \citep{astropy:2013, astropy:2018}. This research also made use of {\sc SNCosmo} to manage SEDs of simulated kNe \citep{Barbary2014}.

The contributions from the authors are listed below: {\bf C.N.S.}: led the project; conceived methodology; wrote simulation software; formulated and carried out calculations; obtained, validated, and interpreted results; writing - original draft - editing - final. {\bf R.B.}: formulated calculations; methodology; software; investigation; validation; writing - editing. {\bf H.V.P.}: project conceptualisation; formulated calculations; methodology; investigation; project administration; supervision; validation; writing - editing; funding acquisition. {\bf S.R., O.K.,} and {\bf R.T.W}: provided software to calculate SEDs and light curves for a population of kNe; writing - editing.

This work was performed in part at the Aspen Center for Physics, which is supported by National Science Foundation grant PHY-1607611. This work was also partially supported by a grant from the Simons Foundation. H.V.P. was partially supported by the European Research Council (ERC) under the European Community's Seventh Framework Programme (FP7/2007-2013)/ERC grant agreement number 306478-CosmicDawn. S.R., R.B., and H.V.P. have been supported by the research environment grant ``Gravitational Radiation and Electromagnetic Astrophysical Transients (GREAT)" funded by the Swedish Research Council (VR) under Dnr 2016-06012. S.R. has been supported by the Swedish Research Council (VR) under grant number 2016- 03657\_3 and by the Swedish National Space Board under grant number Dnr. 107/16. O.K. and R.T.W. are supported by the US Department of Energy through the Los Alamos National Laboratory (LANL). LANL is operated by Triad National Security, LLC, for the National Nuclear Security Administration of U.S. Department of Energy (Contract No. 89233218CNA000001). O.K. and R.T.W. are partially funded by the LANL Directed Research Grant 20190021DR, and used resources provided by LANL Institutional Computing.

The DESC acknowledges ongoing support from the Institut National de Physique Nucl\'eaire et de Physique des Particules in France; the Science \& Technology Facilities Council in the United Kingdom; and the Department of Energy, the National Science Foundation, and the LSST Corporation in the United States. DESC uses resources of the IN2P3 Computing Center (CC-IN2P3--Lyon/Villeurbanne - France) funded by the Centre National de la Recherche Scientifique; the National Energy Research Scientific Computing Center, a DOE Office of Science User Facility supported by the Office of Science of the U.S.\ Department of Energy under Contract No.\ DE-AC02-05CH11231; STFC DiRAC HPC Facilities, funded by UK BIS National E-infrastructure capital grants; and the UK particle physics grid, supported by the GridPP Collaboration.



\bibliographystyle{mnras}
\bibliography{kne_paper1_refs_w_jabbrev}

\begin{thebibliography}{}
\makeatletter
\relax
\def\mn@urlcharsother{\let\do\@makeother \do\$\do\&\do\#\do\^\do\_\do\%\do\~}
\def\mn@doi{\begingroup\mn@urlcharsother \@ifnextchar [ {\mn@doi@}
  {\mn@doi@[]}}
\def\mn@doi@[#1]#2{\def\@tempa{#1}\ifx\@tempa\@empty \href
  {http://dx.doi.org/#2} {doi:#2}\else \href {http://dx.doi.org/#2} {#1}\fi
  \endgroup}
\def\mn@eprint#1#2{\mn@eprint@#1:#2::\@nil}
\def\mn@eprint@arXiv#1{\href {http://arxiv.org/abs/#1} {{\tt arXiv:#1}}}
\def\mn@eprint@dblp#1{\href {http://dblp.uni-trier.de/rec/bibtex/#1.xml}
  {dblp:#1}}
\def\mn@eprint@#1:#2:#3:#4\@nil{\def\@tempa {#1}\def\@tempb {#2}\def\@tempc
  {#3}\ifx \@tempc \@empty \let \@tempc \@tempb \let \@tempb \@tempa \fi \ifx
  \@tempb \@empty \def\@tempb {arXiv}\fi \@ifundefined
  {mn@eprint@\@tempb}{\@tempb:\@tempc}{\expandafter \expandafter \csname
  mn@eprint@\@tempb\endcsname \expandafter{\@tempc}}}

\bibitem[\protect\citeauthoryear{Abbott et~al.,}{Abbott
  et~al.}{2016}]{TheLIGOScientificCollaboration2016}
Abbott B.~P.,  et~al., 2016, \mn@doi [ApJ] {10.3847/2041-8205/832/2/L21}, 832,
  L21

\bibitem[\protect\citeauthoryear{Abbott et~al.,}{Abbott
  et~al.}{2017}]{TheLIGOScientificCollaboration2017}
Abbott B.~P.,  et~al., 2017, \mn@doi [Phys. Rev. Lett.]
  {10.1103/PhysRevLett.119.161101}, 119, 161101

\bibitem[\protect\citeauthoryear{Abbott et~al.,}{Abbott
  et~al.}{2018}]{TheKAGRACollaboration2013}
Abbott B.~P.,  et~al., 2018, \mn@doi [Living Rev. in Relativ.]
  {10.1007/s41114-018-0012-9}, 21, 3

\bibitem[\protect\citeauthoryear{Acernese et~al.,}{Acernese
  et~al.}{2007}]{Acernese2007}
Acernese F.,  et~al., 2007, \mn@doi [Classical and Quantum Gravity]
  {10.1088/0264-9381/24/19/S29}, 24, S671

\bibitem[\protect\citeauthoryear{Andreoni et~al.,}{Andreoni
  et~al.}{2018}]{Andreoni2018}
Andreoni I.,  et~al., 2018, preprint (\mn@eprint {arXiv} {1812.03161})

\bibitem[\protect\citeauthoryear{Awan et~al.,}{Awan et~al.}{2016}]{Awan2016}
Awan H.,  et~al., 2016, \mn@doi [ApJ] {10.3847/0004-637X/829/1/50}, 829, 50

\bibitem[\protect\citeauthoryear{Barbary}{Barbary}{2018}]{Barbary2014}
Barbary K.,  2018, sncosmo v1.6.0, \mn@doi{10.5281/ZENODO.11938}, \url
  {https://doi.org/10.5281/zenodo.11938}

\bibitem[\protect\citeauthoryear{Barnes \& Kasen}{Barnes \&
  Kasen}{2013}]{Barnes2013}
Barnes J.,  Kasen D.,  2013, \mn@doi [ApJ] {10.1088/0004-637X/775/1/18}, 775,
  18

\bibitem[\protect\citeauthoryear{Biscans et~al.,}{Biscans
  et~al.}{2018}]{Biscans2018}
Biscans S.,  et~al., 2018, \mn@doi [Classical and Quantum Gravity]
  {10.1088/1361-6382/aaa4aa}, 35, 055004

\bibitem[\protect\citeauthoryear{Blanchard et~al.,}{Blanchard
  et~al.}{2017}]{Blanchard2017}
Blanchard P.~K.,  et~al., 2017, \mn@doi [ApJ] {10.3847/2041-8213/aa9055}, 848,
  L22

\bibitem[\protect\citeauthoryear{Blanchet}{Blanchet}{2014}]{Blanchet2013}
Blanchet L.,  2014, \mn@doi [Living Rev. in Relativ.] {10.12942/lrr-2014-2},
  17, 2

\bibitem[\protect\citeauthoryear{Canton \& Harry}{Canton \&
  Harry}{2017}]{Canton2017}
Canton T.~D.,  Harry I.~W.,  2017, preprint (\mn@eprint {arXiv} {1705.01845})

\bibitem[\protect\citeauthoryear{Chamberlain \& Yunes}{Chamberlain \&
  Yunes}{2017}]{Chamberlain2017}
Chamberlain K.,  Yunes N.,  2017, \mn@doi [Phys. Rev. D]
  {10.1103/PhysRevD.96.084039}, 96, 084039

\bibitem[\protect\citeauthoryear{Chen, Holz, Miller, Evans, Vitale  \&
  Creighton}{Chen et~al.}{2017a}]{Chen2017a}
Chen H.-Y.,  Holz D.~E.,  Miller J.,  Evans M.,  Vitale S.,   Creighton J.,
  2017a, preprint (\mn@eprint {arXiv} {1709.08079})

\bibitem[\protect\citeauthoryear{Chen, Essick, Vitale, Holz  \&
  Katsavounidis}{Chen et~al.}{2017b}]{Chen2017b}
Chen H.-Y.,  Essick R.,  Vitale S.,  Holz D.~E.,   Katsavounidis E.,  2017b,
  \mn@doi [ApJ] {10.3847/1538-4357/835/1/31}, 835, 31

\bibitem[\protect\citeauthoryear{Chen, Fishbach  \& Holz}{Chen
  et~al.}{2018}]{Chen2018}
Chen H.-Y.,  Fishbach M.,   Holz D.~E.,  2018, \mn@doi [Nature]
  {10.1038/s41586-018-0606-0}, 562, 545

\bibitem[\protect\citeauthoryear{Cho \& Lee}{Cho \& Lee}{2018}]{Cho2018}
Cho H.-S.,  Lee C.-H.,  2018, \mn@doi [J. of the Korean Phys. Soc.]
  {10.3938/jkps.72.1}, 72, 1

\bibitem[\protect\citeauthoryear{Connaughton et~al.,}{Connaughton
  et~al.}{2015}]{Connaughton2014}
Connaughton V.,  et~al., 2015, \mn@doi [ApJ Suppl. Series]
  {10.1088/0067-0049/216/2/32}, 216, 32

\bibitem[\protect\citeauthoryear{Coward et~al.,}{Coward
  et~al.}{2011}]{Coward2011}
Coward D.~M.,  et~al., 2011, \mn@doi [MNRAS: Lett.]
  {10.1111/j.1745-3933.2011.01072.x}, 415, L26

\bibitem[\protect\citeauthoryear{Cowperthwaite et~al.,}{Cowperthwaite
  et~al.}{2017}]{Cowperthwaite2017}
Cowperthwaite P.~S.,  et~al., 2017, \mn@doi [ApJ] {10.3847/2041-8213/aa8fc7},
  848, L17

\bibitem[\protect\citeauthoryear{Cowperthwaite, Villar, Scolnic  \&
  Berger}{Cowperthwaite et~al.}{2018}]{Cowperthwaite2018}
Cowperthwaite P.~S.,  Villar V.~A.,  Scolnic D.~M.,   Berger E.,  2018,
  preprint (\mn@eprint {arXiv} {1811.03098})

\bibitem[\protect\citeauthoryear{{Dal Canton} et~al.,}{{Dal Canton}
  et~al.}{2014}]{DalCanton2014}
{Dal Canton} T.,  et~al., 2014, \mn@doi [Phys. Rev. D]
  {10.1103/PhysRevD.90.082004}, 90, 082004

\bibitem[\protect\citeauthoryear{Dalal, Holz, Hughes  \& Jain}{Dalal
  et~al.}{2006}]{Dalal2006}
Dalal N.,  Holz D.~E.,  Hughes S.~A.,   Jain B.,  2006, \mn@doi [Phys. Rev. D]
  {10.1103/PhysRevD.74.063006}, 74, 063006

\bibitem[\protect\citeauthoryear{Davis et~al.,}{Davis et~al.}{2011}]{Davis2010}
Davis T.~M.,  et~al., 2011, \mn@doi [ApJ] {10.1088/0004-637X/741/1/67}, 741, 67

\bibitem[\protect\citeauthoryear{Delgado \& Reuter}{Delgado \&
  Reuter}{2016}]{Delgado2016}
Delgado F.,  Reuter M.~A.,  2016, in Peck A.~B.,  Seaman R.~L.,   Benn C.~R.,
  eds, ~1 Vol. 9910, Observatory Operations: Strategies, Processes, and Systems
  VI. International Society for Optics and Photonics, p. 991013,
  \mn@doi{10.1117/12.2233630}, \url
  {http://proceedings.spiedigitallibrary.org/proceeding.aspx?doi=10.1117/12.2233630}

\bibitem[\protect\citeauthoryear{{Della Valle} et~al.,}{{Della Valle}
  et~al.}{2018}]{DellaValle2018b}
{Della Valle} M.,  et~al., 2018, \mn@doi [Monthly Notices of the Royal
  Astronomical Society] {10.1093/mnras/sty2541}, 481, 4355

\bibitem[\protect\citeauthoryear{{Di Valentino}, Holz, Melchiorri  \&
  Renzi}{{Di Valentino} et~al.}{2018}]{DiValentino2018}
{Di Valentino} E.,  Holz D.~E.,  Melchiorri A.,   Renzi F.,  2018, \mn@doi
  [Phys. Rev. D] {10.1103/PhysRevD.98.083523}, 98, 083523

\bibitem[\protect\citeauthoryear{Fairhurst}{Fairhurst}{2014}]{Fairhurst2014}
Fairhurst S.,  2014, \mn@doi [J. Phys.: Conf. Series]
  {10.1088/1742-6596/484/1/012007}, 484, 012007

\bibitem[\protect\citeauthoryear{Feeney, Peiris, Williamson, Nissanke,
  Mortlock, Alsing  \& Scolnic}{Feeney et~al.}{2018}]{Feeney2018}
Feeney S.~M.,  Peiris H.~V.,  Williamson A.~R.,  Nissanke S.~M.,  Mortlock
  D.~J.,  Alsing J.,   Scolnic D.,  2018, preprint (\mn@eprint {arXiv}
  {1802.03404})

\bibitem[\protect\citeauthoryear{Fern{\'{a}}ndez, Foucart, Kasen, Lippuner,
  Desai  \& Roberts}{Fern{\'{a}}ndez et~al.}{2017}]{Fernandez2016}
Fern{\'{a}}ndez R.,  Foucart F.,  Kasen D.,  Lippuner J.,  Desai D.,   Roberts
  L.~F.,  2017, \mn@doi [Classical and Quantum Gravity]
  {10.1088/1361-6382/aa7a77}, 34, 154001

\bibitem[\protect\citeauthoryear{Foucart et~al.,}{Foucart
  et~al.}{2014}]{Foucart2014}
Foucart F.,  et~al., 2014, \mn@doi [Phys. Rev. D] {10.1103/PhysRevD.90.024026},
  90, 024026

\bibitem[\protect\citeauthoryear{Greiner}{Greiner}{1995}]{Greiner1995}
Greiner J.,  1995, \mn@doi [Ann. of the N.Y. Acad. of Sci.]
  {10.1111/j.1749-6632.1995.tb17579.x}, 759, 429

\bibitem[\protect\citeauthoryear{Grossman, Korobkin, Rosswog  \&
  Piran}{Grossman et~al.}{2014}]{Grossman2014}
Grossman D.,  Korobkin O.,  Rosswog S.,   Piran T.,  2014, \mn@doi [MNRAS]
  {10.1093/mnras/stt2503}, 439, 757

\bibitem[\protect\citeauthoryear{Guillochon, Nicholl, Villar, Mockler, Narayan,
  Mandel, Berger  \& Williams}{Guillochon et~al.}{2018}]{Guillochon2017}
Guillochon J.,  Nicholl M.,  Villar V.~A.,  Mockler B.,  Narayan G.,  Mandel
  K.~S.,  Berger E.,   Williams P. K.~G.,  2018, \mn@doi [ApJ Suppl. Series]
  {10.3847/1538-4365/aab761}, 236, 6

\bibitem[\protect\citeauthoryear{Hinderer et~al.,}{Hinderer
  et~al.}{2018}]{Hinderer2018}
Hinderer T.,  et~al., 2018, preprint (\mn@eprint {arXiv} {1808.03836})

\bibitem[\protect\citeauthoryear{Holz \& Hughes}{Holz \&
  Hughes}{2005}]{Holz2005}
Holz D.~E.,  Hughes S.~A.,  2005, \mn@doi [ApJ] {10.1086/431341}, 629, 15

\bibitem[\protect\citeauthoryear{Huang, Middleton, Ng, Vitale  \& Veitch}{Huang
  et~al.}{2018}]{Huang2018}
Huang Y.,  Middleton H.,  Ng K. K.~Y.,  Vitale S.,   Veitch J.,  2018, preprint
  (\mn@eprint {arXiv} {1810.10035})

\bibitem[\protect\citeauthoryear{Ivezi{\'{c}} et~al.,}{Ivezi{\'{c}}
  et~al.}{2008}]{Ivezic2008}
Ivezi{\'{c}} {\v{Z}}.,  et~al., 2008, preprint (\mn@eprint {arXiv} {0805.2366})

\bibitem[\protect\citeauthoryear{Ivezi{\'{c}}, Jones  \& Ribeiro}{Ivezi{\'{c}}
  et~al.}{2018}]{LSSTwpcall2018}
Ivezi{\'{c}} {\v{Z}}.,  Jones L.,   Ribeiro T.,  2018, {Call for White Papers
  on LSST Cadence Optimization}, \url {http://ls.st/doc-28382}

\bibitem[\protect\citeauthoryear{Jin et~al.,}{Jin et~al.}{2018}]{Jin2018}
Jin Z.-P.,  et~al., 2018, \mn@doi [ApJ] {10.3847/1538-4357/aab76d}, 857, 128

\bibitem[\protect\citeauthoryear{Juric et~al.,}{Juric
  et~al.}{2018}]{LSSTDDP2018}
Juric M.,  et~al., 2018, {Large Synoptic Survey Telescope (LSST) Systems
  Engineering Data Products Definition Document}, \url
  {https://docushare.lsstcorp.org/docushare/dsweb/Get/LSE-163/}

\bibitem[\protect\citeauthoryear{Kasen, Badnell  \& Barnes}{Kasen
  et~al.}{2013}]{Kasen2013}
Kasen D.,  Badnell N.~R.,   Barnes J.,  2013, \mn@doi [ApJ]
  {10.1088/0004-637X/774/1/25}, 774, 25

\bibitem[\protect\citeauthoryear{Kasliwal et~al.,}{Kasliwal
  et~al.}{2017}]{Kasliwal2017}
Kasliwal M.~M.,  et~al., 2017, \mn@doi [Science] {10.1126/science.aap9455},
  358, 1559

\bibitem[\protect\citeauthoryear{Kelley, Mandel  \& Ramirez-Ruiz}{Kelley
  et~al.}{2013}]{Kelley2012}
Kelley L.~Z.,  Mandel I.,   Ramirez-Ruiz E.,  2013, \mn@doi [Phys. Rev. D]
  {10.1103/PhysRevD.87.123004}, 87, 123004

\bibitem[\protect\citeauthoryear{Kessler et~al.,}{Kessler
  et~al.}{2009}]{Kessler2009}
Kessler R.,  et~al., 2009, \mn@doi [PASP] {10.1086/605984}, 121, 1028

\bibitem[\protect\citeauthoryear{Kessler et~al.,}{Kessler
  et~al.}{2015}]{Kessler2015}
Kessler R.,  et~al., 2015, \mn@doi [AJ] {10.1088/0004-6256/150/6/172}, 150, 172

\bibitem[\protect\citeauthoryear{Kochanek \& Piran}{Kochanek \&
  Piran}{1993}]{Kochanek1993}
Kochanek C.~S.,  Piran T.,  1993, \mn@doi [ApJ] {10.1086/187083}, 417, L17

\bibitem[\protect\citeauthoryear{Li \& Paczy{\'{n}}ski}{Li \&
  Paczy{\'{n}}ski}{1998}]{Li1998}
Li L.-X.,  Paczy{\'{n}}ski B.,  1998, \mn@doi [ApJ] {10.1086/311680}, 507, L59

\bibitem[\protect\citeauthoryear{Lochner et~al.,}{Lochner
  et~al.}{2018}]{Lochner2018a}
Lochner M.,  et~al., 2018, preprint (\mn@eprint {arXiv} {1812.00515})

\bibitem[\protect\citeauthoryear{Margutti et~al.,}{Margutti
  et~al.}{2018}]{Margutti2018}
Margutti R.,  et~al., 2018, preprint (\mn@eprint {arXiv} {1812.04051})

\bibitem[\protect\citeauthoryear{Metzger}{Metzger}{2017}]{Metzger2016}
Metzger B.~D.,  2017, \mn@doi [Living Rev. in Relativ.]
  {10.1007/s41114-017-0006-z}, 20, 3

\bibitem[\protect\citeauthoryear{Metzger \& Berger}{Metzger \&
  Berger}{2012}]{Metzger2012}
Metzger B.~D.,  Berger E.,  2012, \mn@doi [ApJ] {10.1088/0004-637X/746/1/48},
  746

\bibitem[\protect\citeauthoryear{Mortlock, Feeney, Peiris, Williamson  \&
  Nissanke}{Mortlock et~al.}{2018}]{Mortlock2018}
Mortlock D.~J.,  Feeney S.~M.,  Peiris H.~V.,  Williamson A.~R.,   Nissanke
  S.~M.,  2018, preprint (\mn@eprint {arXiv} {1811.11723})

\bibitem[\protect\citeauthoryear{Naghib, Yoachim, Vanderbei, Connolly  \&
  Jones}{Naghib et~al.}{2018}]{Naghib2018}
Naghib E.,  Yoachim P.,  Vanderbei R.~J.,  Connolly A.~J.,   Jones R.~L.,
  2018, preprint (\mn@eprint {arXiv} {1810.04815})

\bibitem[\protect\citeauthoryear{Nissanke, Holz, Hughes, Dalal  \&
  Sievers}{Nissanke et~al.}{2010}]{Nissanke2010}
Nissanke S.,  Holz D.~E.,  Hughes S.~A.,  Dalal N.,   Sievers J.~L.,  2010,
  \mn@doi [ApJ] {10.1088/0004-637X/725/1/496}, 725, 496

\bibitem[\protect\citeauthoryear{Nissanke, Holz, Dalal, Hughes, Sievers  \&
  Hirata}{Nissanke et~al.}{2013}]{Nissanke2013}
Nissanke S.,  Holz D.~E.,  Dalal N.,  Hughes S.~A.,  Sievers J.~L.,   Hirata
  C.~M.,  2013, preprint (\mn@eprint {arXiv} {1307.2638})

\bibitem[\protect\citeauthoryear{Nitz et~al.,}{Nitz et~al.}{2018}]{Nitz2018}
Nitz A.,  et~al., 2018, {gwastro/pycbc: PyCBC v1.13.2 Release},
  \mn@doi{10.5281/zenodo.1596771}, \url {https://zenodo.org/record/1596771}

\bibitem[\protect\citeauthoryear{Perego, Radice  \& Bernuzzi}{Perego
  et~al.}{2017}]{Perego2017}
Perego A.,  Radice D.,   Bernuzzi S.,  2017, \mn@doi [ApJ]
  {10.3847/2041-8213/aa9ab9}, 850, L37

\bibitem[\protect\citeauthoryear{Pinto \& Eastman}{Pinto \&
  Eastman}{2000}]{Pinto2000}
Pinto P.~A.,  Eastman R.~G.,  2000, preprint (\mn@eprint {arXiv} {0006171})

\bibitem[\protect\citeauthoryear{{Planck Collaboration} et~al.,}{{Planck
  Collaboration} et~al.}{2016}]{PlanckCollaboration2015a}
{Planck Collaboration} et~al., 2016, \mn@doi [A{\&}A]
  {10.1051/0004-6361/201525830}, 594, A13

\bibitem[\protect\citeauthoryear{Price-Whelan et~al.,}{Price-Whelan
  et~al.}{2018}]{astropy:2018}
Price-Whelan A.~M.,  et~al., 2018, \mn@doi [AJ] {10.3847/1538-3881/aabc4f},
  156, 123

\bibitem[\protect\citeauthoryear{Reuter, Cook, Delgado, Petry  \&
  Ridgway}{Reuter et~al.}{2016}]{Reuter2016}
Reuter M.~A.,  Cook K.~H.,  Delgado F.,  Petry C.~E.,   Ridgway S.~T.,  2016,
  in Angeli G.~Z.,  Dierickx P.,  eds, ~1 Vol. 9911, Modeling, Systems
  Engineering, and Project Management for Astronomy VI. International Society
  for Optics and Photonics, p. 991125, \mn@doi{10.1117/12.2232680}, \url
  {http://proceedings.spiedigitallibrary.org/proceeding.aspx?doi=10.1117/12.2232680}

\bibitem[\protect\citeauthoryear{Ridgway, Matheson, Mighell, Olsen  \&
  Howell}{Ridgway et~al.}{2014}]{Ridgway2014}
Ridgway S.~T.,  Matheson T.,  Mighell K.~J.,  Olsen K.~A.,   Howell S.~B.,
  2014, \mn@doi [ApJ] {10.1088/0004-637X/796/1/53}, 796, 53

\bibitem[\protect\citeauthoryear{Robitaille et~al.,}{Robitaille
  et~al.}{2013}]{astropy:2013}
Robitaille T.~P.,  et~al., 2013, \mn@doi [A{\&}A]
  {10.1051/0004-6361/201322068}, 558, A33

\bibitem[\protect\citeauthoryear{Rosswog}{Rosswog}{2005}]{Rosswog2005}
Rosswog S.,  2005, \mn@doi [ApJ] {10.1086/497062}, 634, 1202

\bibitem[\protect\citeauthoryear{Rosswog}{Rosswog}{2013}]{Rosswog2013}
Rosswog S.,  2013, \mn@doi [Philos. Trans. of the R. Soc. A: Math., Phys. and
  Eng. Sci.] {10.1098/rsta.2012.0272}, 371, 20120272

\bibitem[\protect\citeauthoryear{Rosswog}{Rosswog}{2015}]{Rosswog2015}
Rosswog S.,  2015, \mn@doi [Inter. J. of Modern Phys. D]
  {10.1142/S0218271815300128}, 24, 1530012

\bibitem[\protect\citeauthoryear{Rosswog, Piran  \& Nakar}{Rosswog
  et~al.}{2013}]{Rosswog2012}
Rosswog S.,  Piran T.,   Nakar E.,  2013, \mn@doi [MNRAS]
  {10.1093/mnras/sts708}, 430, 2585

\bibitem[\protect\citeauthoryear{Rosswog, Feindt, Korobkin, Wu, Sollerman,
  Goobar  \& Martinez-Pinedo}{Rosswog et~al.}{2017}]{Rosswog2016a}
Rosswog S.,  Feindt U.,  Korobkin O.,  Wu M.-R.,  Sollerman J.,  Goobar A.,
  Martinez-Pinedo G.,  2017, \mn@doi [Classical and Quantum Gravity]
  {10.1088/1361-6382/aa68a9}, 34, 104001

\bibitem[\protect\citeauthoryear{Rosswog, Sollerman, Feindt, Goobar, Korobkin,
  Wollaeger, Fremling  \& Kasliwal}{Rosswog et~al.}{2018}]{Rosswog2018}
Rosswog S.,  Sollerman J.,  Feindt U.,  Goobar A.,  Korobkin O.,  Wollaeger R.,
   Fremling C.,   Kasliwal M.~M.,  2018, \mn@doi [A{\&}A]
  {10.1051/0004-6361/201732117}, 615, A132

\bibitem[\protect\citeauthoryear{Schlafly \& Finkbeiner}{Schlafly \&
  Finkbeiner}{2011}]{Schlafly2011}
Schlafly E.~F.,  Finkbeiner D.~P.,  2011, \mn@doi [ApJ]
  {10.1088/0004-637X/737/2/103}, 737, 103

\bibitem[\protect\citeauthoryear{Schutz}{Schutz}{1986}]{Schutz1986a}
Schutz B.~F.,  1986, \mn@doi [Nature] {10.1038/323310a0}, 323, 310

\bibitem[\protect\citeauthoryear{Schutz}{Schutz}{2011}]{Schutz2011}
Schutz B.~F.,  2011, \mn@doi [Classical and Quantum Gravity]
  {10.1088/0264-9381/28/12/125023}, 28, 125023

\bibitem[\protect\citeauthoryear{Scolnic et~al.,}{Scolnic
  et~al.}{2017}]{Scolnic2017a}
Scolnic D.,  et~al., 2017, \mn@doi [ApJ] {10.3847/2041-8213/aa9d82}, 852, L3

\bibitem[\protect\citeauthoryear{Scolnic et~al.,}{Scolnic
  et~al.}{2018}]{Scolnic2018}
Scolnic D.~M.,  et~al., 2018, preprint (\mn@eprint {arXiv} {1812.00516})

\bibitem[\protect\citeauthoryear{Smartt et~al.,}{Smartt
  et~al.}{2017}]{Smartt2017}
Smartt S.~J.,  et~al., 2017, \mn@doi [Nature] {10.1038/nature24303}

\bibitem[\protect\citeauthoryear{Soares-Santos et~al.,}{Soares-Santos
  et~al.}{2017}]{Soares-Santos2017}
Soares-Santos M.,  et~al., 2017, \mn@doi [ApJ] {10.3847/2041-8213/aa9059}, 848,
  L16

\bibitem[\protect\citeauthoryear{Tanaka \& Hotokezaka}{Tanaka \&
  Hotokezaka}{2013}]{Tanaka2013}
Tanaka M.,  Hotokezaka K.,  2013, \mn@doi [ApJ] {10.1088/0004-637X/775/2/113},
  775, 113

\bibitem[\protect\citeauthoryear{Tanaka et~al.,}{Tanaka
  et~al.}{2017}]{Tanaka2017}
Tanaka M.,  et~al., 2017, \mn@doi [PASJ] {10.1093/pasj/psx121}, 69

\bibitem[\protect\citeauthoryear{Tanvir et~al.,}{Tanvir
  et~al.}{2017}]{Tanvir2017}
Tanvir N.~R.,  et~al., 2017, \mn@doi [ApJ] {10.3847/2041-8213/aa90b6}, 848, L27

\bibitem[\protect\citeauthoryear{{The LIGO Scientific Collaboration}
  et~al.,}{{The LIGO Scientific Collaboration} et~al.}{2017a}]{Abbott2017}
{The LIGO Scientific Collaboration} et~al., 2017a, \mn@doi [Nature]
  {10.1038/nature24471}, 551, 85

\bibitem[\protect\citeauthoryear{{The LIGO Scientific Collaboration}
  et~al.,}{{The LIGO Scientific Collaboration} et~al.}{2017b}]{Abbott2017a}
{The LIGO Scientific Collaboration} et~al., 2017b, \mn@doi [ApJ]
  {10.3847/2041-8213/aa920c}, 848, L13

\bibitem[\protect\citeauthoryear{{The LIGO Scientific Collaboration}
  et~al.,}{{The LIGO Scientific Collaboration} et~al.}{2017c}]{Abbott2017b}
{The LIGO Scientific Collaboration} et~al., 2017c, \mn@doi [ApJ]
  {10.3847/2041-8213/aa920c}, 848, L13

\bibitem[\protect\citeauthoryear{{The LIGO Scientific Collaboration}
  et~al.,}{{The LIGO Scientific Collaboration}
  et~al.}{2017d}]{Collaboration2017}
{The LIGO Scientific Collaboration} et~al., 2017d, \mn@doi [ApJ Lett.]
  {10.3847/2041-8213/aa91c9}, 848, 59

\bibitem[\protect\citeauthoryear{{The LIGO Scientific Collaboration}
  et~al.,}{{The LIGO Scientific Collaboration}
  et~al.}{2018a}]{TheLIGOScientificCollaboration2018}
{The LIGO Scientific Collaboration} et~al., 2018a, preprint (\mn@eprint {arXiv}
  {1811.12907})

\bibitem[\protect\citeauthoryear{{The LIGO Scientific Collaboration}
  et~al.,}{{The LIGO Scientific Collaboration} et~al.}{2018b}]{Abbott2018}
{The LIGO Scientific Collaboration} et~al., 2018b, \mn@doi [Living Rev. in
  Relativ.] {10.1007/s41114-018-0012-9}, 21, 3

\bibitem[\protect\citeauthoryear{{The LSST Dark Energy Science Collaboration}
  et~al.,}{{The LSST Dark Energy Science Collaboration}
  et~al.}{2018}]{TheLSSTDarkEnergyScienceCollaboration2018}
{The LSST Dark Energy Science Collaboration} et~al., 2018, preprint (\mn@eprint
  {arXiv} {1809.01669})

\bibitem[\protect\citeauthoryear{{The LSST Science Collaboration} et~al.,}{{The
  LSST Science Collaboration} et~al.}{2009}]{LSSTScienceCollaboration2009a}
{The LSST Science Collaboration} et~al., 2009, preprint (\mn@eprint {arXiv}
  {0912.0201})

\bibitem[\protect\citeauthoryear{{The LSST Science Collaboration} et~al.,}{{The
  LSST Science Collaboration} et~al.}{2017}]{LSSTScienceCollaboration2017}
{The LSST Science Collaboration} et~al., 2017, preprint (\mn@eprint {arXiv}
  {1708.04058})

\bibitem[\protect\citeauthoryear{Usman et~al.,}{Usman et~al.}{2016}]{Usman2016}
Usman S.~A.,  et~al., 2016, \mn@doi [Classical and Quantum Gravity]
  {10.1088/0264-9381/33/21/215004}, 33, 215004

\bibitem[\protect\citeauthoryear{Vangioni, Goriely, Daigne, Fran{\c{c}}ois  \&
  Belczynski}{Vangioni et~al.}{2016}]{Vangioni2016}
Vangioni E.,  Goriely S.,  Daigne F.,  Fran{\c{c}}ois P.,   Belczynski K.,
  2016, \mn@doi [MNRAS] {10.1093/mnras/stv2296}, 455, 17

\bibitem[\protect\citeauthoryear{Villar, Berger, Metzger  \& Guillochon}{Villar
  et~al.}{2017a}]{Villar2017a}
Villar V.~A.,  Berger E.,  Metzger B.~D.,   Guillochon J.,  2017a, \mn@doi
  [ApJ] {10.3847/1538-4357/aa8fcb}, 849, 70

\bibitem[\protect\citeauthoryear{Villar et~al.,}{Villar
  et~al.}{2017b}]{Villar2017b}
Villar V.~A.,  et~al., 2017b, \mn@doi [ApJ] {10.3847/2041-8213/aa9c84}, 851,
  L21

\bibitem[\protect\citeauthoryear{Vitale \& Chen}{Vitale \&
  Chen}{2018}]{Vitale2018}
Vitale S.,  Chen H.-Y.,  2018, \mn@doi [Phys. Rev. Lett.]
  {10.1103/PhysRevLett.121.021303}, 121, 021303

\bibitem[\protect\citeauthoryear{Wollaeger, van Rossum, Graziani, Couch,
  {Jordan IV}, Lamb  \& Moses}{Wollaeger et~al.}{2013}]{Wollaeger2013}
Wollaeger R.~T.,  van Rossum D.~R.,  Graziani C.,  Couch S.~M.,  {Jordan IV}
  G.~C.,  Lamb D.~Q.,   Moses G.~A.,  2013, \mn@doi [ApJ Suppl. Series]
  {10.1088/0067-0049/209/2/36}, 209, 36

\bibitem[\protect\citeauthoryear{Wollaeger et~al.,}{Wollaeger
  et~al.}{2018}]{Wollaeger2018}
Wollaeger R.~T.,  et~al., 2018, \mn@doi [MNRAS] {10.1093/mnras/sty1018}, 478,
  3298

\makeatother
\end{thebibliography}



\appendix
\section{Tabulated Detection Results}\label{sec:appendix}
Here we present tabulated results for the total number of serendipitously detected kNe over the ten-year planned survey duration for each kN model, each survey strategy, and each set of detection criteria. This separately tabulates the subset of detected kNe that are below the detection threshold of the ALV GW observatories according to our calculations outlined in Sec.\ \ref{subsec:GWsignals}.

Additionally, for comparison with other works, we include the detection criteria used by \citet{Cowperthwaite2018}:
\begin{itemize}
  \item A minimum of three measurements with $\mathrm{SNR} \geq 5$;
  \item At least two measurements with $\mathrm{SNR} \geq 5$ in the same filter.
\end{itemize}

These criteria are part of a set of criteria identifying light curves observed with LSST that have been detected (the criteria above), exhibit a rise, and have colour information \citep{Cowperthwaite2018}; see \citet{Cowperthwaite2018} for the `rise' and `colour' criteria. These criteria were developed with the goal of assessing the adequacy of LSST-only light curves for determining astrophysical parameters. As \citet{Cowperthwaite2018} shows, the WFD survey of LSST will not obtain many kNe light curves that are useful for inferring such astrophysical parameters. While our science case is different, we consider the detection criteria for comparison.

These detection criteria impose fewer explicit restrictions on the light curve data than that of \citet{Scolnic2017a}. However, these criteria require a light curve with, at minimum, three high-quality data points. Considering the fast evolution of kNe, and the approximate cadence of three days to re-observe a location on the sky \citep{LSSTScienceCollaboration2009a}, this is likely more limiting. Additionally, three light curve measurements should inherently reject asteroids and should also provide shape and/or colour information to reject other types of astrophysical transients. However, these detection criteria do not explicitly enforce such requirements, like the criteria of \citet{Scolnic2017a}.

We label the detection results from these criteria evaluated on coadded observations C1. The same detection criteria evaluated on individual observations we refer to as C2. For the detection criteria described in this section, using a rate of $1500 \, \mathrm{Gpc^{-3} \,yr^{-1}}$, \citet{Cowperthwaite2018} report a rate of $3{-}6$ detected kNe per year for the LSST WFD. For our simulations of the Population BNS kNe, with the lower assumed rate of $1000 \, \mathrm{Gpc^{-3} \,yr^{-1}}$, we find a detection rate of $2{-}5$ kNe per year for C1 and, for C2, a detection rate of $5{-}9$ kNe per year.

\begin{table}
\centering
\resizebox{\columnwidth}{!}{%
  \begin{tabular}{c|c|c|c}
    & & & Number below GW \\
   Survey Strategy & Det. Criteria & Number of Detections & detection threshold \\
  \hline
  opsim\_baseline & S1 & 72 $ \pm $  8 & 42 $ \pm $  6 \\
 & S2 & 58 $ \pm $  7 & 30 $ \pm $  5 \\
 & C1 & 47 $ \pm $  6 & 21 $ \pm $  4 \\
 & C2 & 157 $ \pm $ 12 & 97 $ \pm $  9 \\

  \hline
  alt\_sched\_rolling & S1 & 127 $ \pm $ 11 & 81 $ \pm $  9 \\
 & S2 & 131 $ \pm $ 11 & 84 $ \pm $  9 \\
 & C1 & 90 $ \pm $  9 & 45 $ \pm $  6 \\
 & C2 & 104 $ \pm $ 10 & 56 $ \pm $  7 \\

  \hline
  opsim\_single\_exp & S1 & 84 $ \pm $  9 & 52 $ \pm $  7 \\
 & S2 & 65 $ \pm $  8 & 36 $ \pm $  6 \\
 & C1 & 57 $ \pm $  7 & 27 $ \pm $  5 \\
 & C2 & 170 $ \pm $ 13 & 113 $ \pm $ 10 \\

  \hline
  opsim\_large\_rolling\_3yr & S1 & 65 $ \pm $  8 & 33 $ \pm $  5 \\
 & S2 & 52 $ \pm $  7 & 20 $ \pm $  4 \\
 & C1 & 55 $ \pm $  7 & 22 $ \pm $  4 \\
 & C2 & 136 $ \pm $ 11 & 84 $ \pm $  9 \\

  \hline
  opsim\_large & S1 & 65 $ \pm $  8 & 31 $ \pm $  5 \\
 & S2 & 54 $ \pm $  7 & 22 $ \pm $  4 \\
 & C1 & 45 $ \pm $  6 & 17 $ \pm $  4 \\
 & C2 & 133 $ \pm $ 11 & 77 $ \pm $  8 \\

  \hline
  opsim\_20\_exp & S1 & 117 $ \pm $ 10 & 63 $ \pm $  7 \\
 & S2 & 96 $ \pm $  9 & 48 $ \pm $  6 \\
 & C1 & 72 $ \pm $  8 & 32 $ \pm $  5 \\
 & C2 & 182 $ \pm $ 13 & 116 $ \pm $ 10 \\

  \hline
  fbs\_mixed\_filter\_pairs & S1 & 250 $ \pm $ 15 & 196 $ \pm $ 14 \\
 & S2 & 254 $ \pm $ 15 & 202 $ \pm $ 14 \\
 & C1 & 98 $ \pm $  9 & 54 $ \pm $  7 \\
 & C2 & 142 $ \pm $ 11 & 92 $ \pm $  9 \\

  \end{tabular}
}
 \caption{All detection results for the Single kN model.}\label{tab:all_results1}
\end{table}

\begin{table}
\centering
\resizebox{\columnwidth}{!}{%
  \begin{tabular}{c|c|c|c}
     & & & Number below GW \\
    Survey Strategy & Det. Criteria & Number of Detections & detection threshold \\
    \hline
  opsim\_baseline & S1 & 37 $ \pm $  6 & 18 $ \pm $  4 \\
 & S2 & 27 $ \pm $  5 & 12 $ \pm $  3 \\
 & C1 & 24 $ \pm $  4 & 15 $ \pm $  3 \\
 & C2 & 65 $ \pm $  8 & 40 $ \pm $  6 \\

  \hline
  alt\_sched\_rolling & S1 & 58 $ \pm $  7 & 32 $ \pm $  5 \\
 & S2 & 57 $ \pm $  7 & 31 $ \pm $  5 \\
 & C1 & 43 $ \pm $  6 & 18 $ \pm $  4 \\
 & C2 & 49 $ \pm $  7 & 23 $ \pm $  4 \\

  \hline
  opsim\_single\_exp & S1 & 34 $ \pm $  5 & 20 $ \pm $  4 \\
 & S2 & 25 $ \pm $  5 & 13 $ \pm $  3 \\
 & C1 & 32 $ \pm $  5 & 16 $ \pm $  4 \\
 & C2 & 60 $ \pm $  7 & 39 $ \pm $  6 \\

  \hline
  opsim\_large\_rolling\_3yr & S1 & 38 $ \pm $  6 & 21 $ \pm $  4 \\
 & S2 & 32 $ \pm $  5 & 15 $ \pm $  3 \\
 & C1 & 31 $ \pm $  5 & 17 $ \pm $  4 \\
 & C2 & 58 $ \pm $  7 & 31 $ \pm $  5 \\

  \hline
  opsim\_large & S1 & 22 $ \pm $  4 & 4 $ \pm $  2 \\
 & S2 & 24 $ \pm $  4 & 6 $ \pm $  2 \\
 & C1 & 23 $ \pm $  4 & 12 $ \pm $  3 \\
 & C2 & 55 $ \pm $  7 & 29 $ \pm $  5 \\

  \hline
  opsim\_20\_exp & S1 & 61 $ \pm $  8 & 33 $ \pm $  6 \\
 & S2 & 56 $ \pm $  7 & 25 $ \pm $  5 \\
 & C1 & 31 $ \pm $  6 & 16 $ \pm $  4 \\
 & C2 & 86 $ \pm $  9 & 47 $ \pm $  7 \\

  \hline
  fbs\_mixed\_filter\_pairs & S1 & 85 $ \pm $  9 & 60 $ \pm $  7 \\
 & S2 & 82 $ \pm $  9 & 56 $ \pm $  7 \\
 & C1 & 49 $ \pm $  7 & 28 $ \pm $  5 \\
 & C2 & 65 $ \pm $  8 & 40 $ \pm $  6 \\

\end{tabular}
  }
 \caption{All detection results for the Population BNS model.}\label{tab:all_results2}
\end{table}

\begin{table}
\centering
\resizebox{\columnwidth}{!}{%
  \begin{tabular}{c|c|c}
    Survey Strategy & Det. Criteria & Number of Detections \\
    \hline
  opsim\_baseline & S1 & 1 $ \pm $  1 \\
 & S2 & 1 $ \pm $  1 \\
 & C1 & 0 $ \pm $  0 \\
 & C2 & 1 $ \pm $  1 \\

  \hline
  alt\_sched\_rolling & S1 & 1 $ \pm $  1 \\
 & S2 & 1 $ \pm $  1 \\
 & C1 & 1 $ \pm $  1 \\
 & C2 & 1 $ \pm $  1 \\

  \hline
  opsim\_single\_exp & S1 & 1 $ \pm $  1 \\
 & S2 & 0 $ \pm $  0 \\
 & C1 & 1 $ \pm $  1 \\
 & C2 & 2 $ \pm $  1 \\

  \hline
  opsim\_large\_rolling\_3yr & S1 & 2 $ \pm $  1 \\
 & S2 & 1 $ \pm $  1 \\
 & C1 & 2 $ \pm $  1 \\
 & C2 & 6 $ \pm $  2 \\

  \hline
  opsim\_large & S1 & 0 $ \pm $  0 \\
 & S2 & 0 $ \pm $  0 \\
 & C1 & 0 $ \pm $  0 \\
 & C2 & 2 $ \pm $  1 \\

  \hline
  opsim\_20\_exp & S1 & 0 $ \pm $  0 \\
 & S2 & 0 $ \pm $  0 \\
 & C1 & 1 $ \pm $  1 \\
 & C2 & 2 $ \pm $  1 \\

  \hline
  fbs\_mixed\_filter\_pairs & S1 & 4 $ \pm $  2 \\
 & S2 & 4 $ \pm $  2 \\
 & C1 & 0 $ \pm $  0 \\
 & C2 & 0 $ \pm $  0 \\

\end{tabular}
  }
 \caption{All detection results for the Population NSBH model.}\label{tab:all_results3}
\end{table}


\bsp	
\label{lastpage}
\end{document}